\documentclass[10pt,twocolumn,preprintnumbers,prd,superscriptaddress,nofootinbib,floatfix,showpacs,showkeys,aps]{revtex4-2}

\usepackage{float}
\usepackage{graphicx}  
\usepackage{multirow}
\usepackage{natbib}
\usepackage{pifont}
\usepackage{amssymb}
\usepackage{bm}        
\usepackage{amsfonts}  
\usepackage{amsmath}   
\usepackage{amssymb}   
\usepackage[titletoc]{appendix}
\usepackage{color}
\usepackage{hyperref}
\usepackage{cleveref}
\usepackage[english]{babel}
\usepackage{booktabs}
\usepackage{mathtools}
\usepackage{subfigure}
\usepackage{enumerate}
\usepackage{soul}
\usepackage{tensor}
\DeclareMathOperator{\sgn}{sgn}

\definecolor{vividviolet}{rgb}{0.62, 0.0, 1.0}
\definecolor{amaranth}{rgb}{0.9, 0.17, 0.31}
\definecolor{palatinateblue}{rgb}{0.15, 0.23, 0.89}
\definecolor{brightpink}{rgb}{1.0, 0.0, 0.5}
\definecolor{cornflowerblue}{rgb}{0.39, 0.58, 0.93}
\definecolor{deepcarminepink}{rgb}{0.94, 0.19, 0.22}
\definecolor{radicalred}{rgb}{1.0, 0.21, 0.37}

\hypersetup{ linktoc=all,
    colorlinks, linkcolor={palatinateblue},
    citecolor={brightpink}, urlcolor={amaranth}
}

\begin{document}

\title{Scalar perturbations in two-scalar-field warm inflation}

\author{Orlando Luongo}
\email{orlando.luongo@unicam.it}
\affiliation{University of Camerino, Via Madonna delle Carceri, Camerino, 62032, Italy}
\affiliation{Istituto Nazionale di Fisica Nucleare (INFN), Sezione di Perugia, Perugia, 06123, Italy}
\affiliation{Department of Mathematics and Physics, SUNY Polytechnic Institute, Utica, NY 13502, USA}
\affiliation{INAF - Osservatorio Astronomico di Brera, Milano, Italy}
\affiliation{Al-Farabi Kazakh National University, Almaty, 050040, Kazakhstan}

\author{Tommaso Mengoni}
\email{tommaso.mengoni@unicam.it}
\affiliation{University of Camerino, Via Madonna delle Carceri, Camerino, 62032, Italy}
\affiliation{Istituto Nazionale di Fisica Nucleare (INFN), Sezione di Perugia, Perugia, 06123, Italy}
\affiliation{INAF - Osservatorio Astronomico di Brera, Milano, Italy}

\author{Paulo M. S\'a}
\email{pmsa@ualg.pt}
\affiliation{Departamento de F\'\i sica, Faculdade de Ci\^encias e Tecnologia, Universidade do Algarve, Campus de Gambelas, 8005-139 Faro, Portugal}
\affiliation{Centro de Investiga\c{c}\~ao e Desenvolvimento em Matem\'atica e Aplica\c{c}\~oes (CIDMA), Universidade de Aveiro, Campus de Santiago, 3810-193 Aveiro, Portugal}

\begin{abstract} 
We investigate scalar perturbations in a two-scalar-field cosmological model unifying warm inflation, dark matter, and dark energy.
The inflaton is noncanonically coupled to a second scalar field that acts as dark energy at late times.
Assuming negligible dissipation and perturbations of the dark-energy field, we derive the linear scalar perturbations in a gauge-ready form, including quantum and thermal noise in the inflaton sector and stochastic fluctuations of the energy transfer to radiation.
We numerically solve the coupled perturbation system using a Fokker-Planck formalism and compute the scalar spectral index $n_s$ and the tensor-to-scalar ratio $r$ for dissipation coefficients scaling as $T^p$ with $p=-1,\, 1$ and $3$.
For $p=-1$, we find scenarios compatible with current CMB observations in both the weak and strong dissipative regimes, while the explored scenarios with $p=1$ and $3$ are outside the observationally supported regions.
Compared with standard single-field warm inflation, the two-scalar-field model yields a systematically larger spectral index for a given tensor-to-scalar ratio.
\end{abstract}


\maketitle 
\tableofcontents


\section{Introduction\label{sec:introd}}

Since its inception more than four decades ago, the inflationary paradigm, in which the very early Universe undergoes a phase of quasi-exponential expansion, has become an integral part of the standard cosmological model \cite{Starobinsky:1980te, Guth:1980zm, Linde:1981mu, Albrecht:1982wi, Linde:1983gd}.
Inflation was originally proposed to resolve several shortcomings of the hot Big Bang model, most notably the horizon problem, the flatness problem, and the overabundance of relics, such as magnetic monopoles, predicted by grand unified theories.
A sufficiently long inflationary phase, lasting at least $50$--$60$ $e$-folds, dynamically drives the Universe toward homogeneity, isotropy, and spatial flatness, while diluting any unwanted relics to unobservable levels.

Beyond addressing the initial-condition problems of the hot Big Bang model, inflation provides a compelling mechanism for the origin of cosmic structure: quantum fluctuations of the inflaton, the scalar field driving inflation, seed an almost scale-invariant spectrum of adiabatic, nearly Gaussian curvature perturbations \cite{Mukhanov:1981xt, Starobinsky:1982ee, Guth:1982ec, Bardeen:1983qw}, accompanied by primordial gravitational waves \cite{Starobinsky:1979ty}.

The predictions for the scalar sector have been confirmed to remarkable precision by observations of the cosmic microwave background (CMB) anisotropies: Planck measured the scalar spectral index $n_s = 0.9649 \pm 0.0042$ ($68\%$ C.L.) \cite{Planck:2018jri}, a value that shifts to $n_s = 0.9743 \pm 0.0034$ when data from the Atacama Cosmology Telescope (ACT), CMB lensing, and baryon acoustic oscillations (BAO) from the Dark Energy Spectroscopic Instrument (DESI) are added \cite{ACT:2025fju, ACT:2025tim}.
Tensor modes remain undetected; the combination of Planck, BICEP/Keck, and BAO data yields an upper bound on the tensor-to-scalar ratio of $r<0.032$ ($95\%$ C.L.) at the pivot scale $k_\ast=0.05\,\mathrm{Mpc}^{-1}$ \cite{BICEP:2021xfz, Tristram:2021tvh}.
Together, these constraints favor inflationary models with a small $r$ and a slightly red-tilted spectrum, disfavoring, in standard (cold) inflation, simple potentials such as the quadratic and quartic ones \cite{Planck:2018jri, ACT:2025tim}.

The warm inflation (WI) scenario \cite{Berera:1995ie} provides an alternative to standard cold inflation (CI); for reviews, see Refs.~\cite{Berera:2008ar, Kamali:2023lzq, Berera:2023liv}.
In WI, the inflaton remains coupled to other degrees of freedom throughout inflation, continuously dissipating energy into a radiation bath.
This process is described by a dissipation coefficient $\Upsilon$, which introduces an additional friction term in the inflaton equation of motion; its strength relative to Hubble damping is the dissipation ratio $Q = \Upsilon/(3H)$, with $H$ the Hubble parameter.
As a result, the radiation energy density remains non-negligible and the Universe stays at a temperature $T > H$, with two main consequences: radiation can smoothly take over at the end of inflation, so a distinct reheating phase is not required; and primordial density perturbations are sourced predominantly by thermal rather than quantum fluctuations of the inflaton, an effect that modifies the scalar spectrum but not the tensor one.

Constructing concrete realizations of WI from first principles proved challenging, since large thermal corrections to the inflaton mass tend to spoil the flatness of the potential.
This difficulty was overcome by mechanisms in which the inflaton couples to the thermal bath only indirectly, through heavy intermediate fields \cite{Berera:2002sp, Bastero-Gil:2010dgy, Bastero-Gil:2012akf}, and, more recently, by models in which the inflaton is protected by a symmetry, namely, the Warm Little Inflaton \cite{Bastero-Gil:2016qru} and its extensions \cite{Rosa:2018iwu, Bastero-Gil:2019gao}, and minimal WI, with an axion-like inflaton coupled to non-Abelian gauge fields \cite{Berghaus:2019whh}.
These models yield $\Upsilon \propto T^{p}$, typically with $p = -1$, $1$, or $3$.

WI has been shown to be in good agreement with CMB data \cite{Bartrum:2013fia, Benetti:2016jhf, Bastero-Gil:2017wwl, Montefalcone:2023pvh}, including the latest ACT results \cite{Berera:2025vsu}.
Since thermal effects enhance the scalar power spectrum, the tensor-to-scalar ratio is typically suppressed relative to CI, thereby allowing the simple potentials that are disfavored in CI to be consistent with current observations \cite{Bartrum:2013fia, Benetti:2016jhf, Ballesteros:2023dno, Berera:2025vsu}. 
WI in the strong dissipative regime also offers a way to reconcile inflation with the de Sitter swampland conjectures \cite{Motaharfar:2018zyb, Das:2018hqy, Bertolami:2022knd}.

These features make WI more than a phenomenological alternative to CI: it embeds inflation in a particle-physics framework in which the inflaton interacts with other fields during inflation, as expected given its microphysical origin.
Further investigation of WI is timely, as forthcoming CMB polarization experiments such as the Simons Observatory \cite{SimonsObservatory:2018koc} and LiteBIRD \cite{LiteBIRD:2022cnt} will probe $r \sim 10^{-3}$, testing a large class of inflationary models, both cold and warm.

Remarkably, accelerated expansion is not exclusive to the primordial Universe: type Ia supernovae revealed that the expansion is currently accelerating \cite{Riess:1998cb, Perlmutter:1998np}, a behavior attributed to a dark-energy component that dominates the current energy budget.
In the standard $\Lambda$CDM model, dark energy is identified with a cosmological constant, a description that, although economical and phenomenologically successful, suffers from well-known fine-tuning problems \cite{Weinberg:1988cp}.
Moreover, recent BAO measurements by DESI, combined with CMB and supernova data, show a preference for dynamical dark energy \cite{DESI:2024mwx, DESI:2025zgx}, which can be modeled by a slowly rolling scalar field, known as quintessence \cite{Ratra:1987rm, Wetterich:1987fm, Caldwell:1997ii}.

Since the inflaton is itself a scalar field, it is tempting to regard both periods of accelerated expansion as manifestations of a common underlying dynamics, 
an idea that becomes even more compelling once dark matter is brought into the picture, given that a scalar field oscillating rapidly about the minimum of a quadratic potential behaves, on average, like a pressureless fluid \cite{Turner:1983he}.

Proposals along these lines appeared shortly after the discovery of cosmic acceleration, first as quintessential inflation, where a single field drives both inflation and the present accelerated expansion \cite{Peebles:1998qn}, and then as models unifying inflation, dark energy, and dark matter \cite{Panotopoulos:2007ri, Liddle:2008bm, Henriques:2009hv, Paliathanasis:2025mvy}.
WI provides a particularly suitable setting for such unification: since radiation is continuously produced during inflation, no reheating phase is needed; the inflaton does not have to decay completely, and its remnant can account for the dark matter.
This idea was already implemented in Ref.~\cite{Henriques:2009hv}, within a two-scalar-field model based on the Salam--Sezgin six-dimensional supergravity theory, and has since been explored in other WI-based scenarios \cite{Rosa:2018iwu, Sa:2020fvn, Zhang:2021ppy, Mishra:2026ktu}.

In the two-scalar-field cosmological model of Ref.~\cite{Sa:2020fvn}, which we study in this paper, the inflaton drives a period of WI and, once dissipation has become negligible, accounts for the cold dark matter.
A second scalar field, coupled to the inflaton through exponential factors multiplying the inflaton's kinetic and potential terms, plays the role of dark energy; at late times, the model reduces to a coupled quintessence scenario \cite{Amendola:1999er}.
The background evolution qualitatively follows the standard cosmological history: an inflationary epoch is succeeded by radiation-, matter-, and dark-energy-dominated eras, each of appropriate duration \cite{Sa:2020fvn, Sa:2021eft}.
The coupling between the two fields has been constrained using low-redshift observational data \cite{DAgostino:2021vvv}.

The gravitational-wave background predicted by this two-scalar-field model has also been investigated.
Using the formalism of continuous Bogoliubov coefficients,\footnote{The formalism of continuous Bogoliubov coefficients was extended to gravitons in Refs.~\cite{Henriques:1993km, Moorhouse:1994nc, Mendes:1994ai} and has since been applied to the computation of gravitational-wave spectra in a variety of cosmological models \cite{Henriques:2009hv, Sa:2007pc, Sa:2008yq, Sa:2010qw, Sa:2011rm, Bouhmadi-Lopez:2009uca, Bouhmadi-Lopez:2012piq, Luongo:2025ovo, Luongo:2026wdr}.} Ref.~\cite{Luongo:2025ovo} computed the full energy spectrum and showed that, in the strong dissipative regime, the signal could be within reach of next-generation detectors \cite{Luongo:2025ovo}. 
The weak dissipative regime was found to further improve the prospects for detection \cite{Luongo:2026wdr}, and the spectrum was shown to be largely insensitive to the initial value of the dark-energy field \cite{Mengoni:2026bog}.

A complete assessment of the model’s viability, however, requires the computation of scalar perturbations, which determine the amplitude and tilt of the primordial curvature power spectrum and, together with the tensor spectrum, the tensor-to-scalar ratio.
This is the aim of the present work.

Scalar perturbations in two-scalar-field WI have been investigated previously \cite{Wang:2018cev, Li:2019zbk, Wang:2019ozs}, but only for models with two canonically normalized inflatons, both of which dissipate into the radiation bath, with no kinetic coupling between them, purely thermal noise in the stochastic perturbation equations, and dissipation coefficients that are either independent of the temperature \cite{Wang:2018cev, Li:2019zbk} or linear in it \cite{Wang:2019ozs}.
A two-scalar-field WI model with a kinetic coupling between the fields, motivated by heterotic string theory, has recently been proposed \cite{Berera:2025jbj}, but its perturbations were estimated only in the effectively single-field regime, with a full treatment deferred to future work.
To the best of our knowledge, the present work provides the first computation of scalar perturbations in two-scalar-field WI with a kinetic coupling between the fields, and goes beyond previous studies in several other respects.
We derive the perturbation equations in a general gauge, for a model in which only the inflaton dissipates, while the second field is not an additional inflaton but accounts for dark energy at late times.
The inflaton perturbation equation includes both quantum and thermal noise, with a dissipation coefficient that scales as $T^{p}$ during inflation and is exponentially suppressed afterward.
In addition, stochastic fluctuations of the dissipative energy transfer are considered in the radiation energy-density perturbation equation.

The paper is organized as follows.
In Sec.~\ref{sec:background}, we review the two-scalar-field cosmological model and describe the background evolution from inflation to the present for a representative choice of parameters and initial conditions.
In Sec.~\ref{sec:perturbation}, starting from the most general scalar-perturbed Friedmann--Lema\^itre--Robertson--Walker (FLRW) metric, we derive the scalar perturbation equations.
In Sec.~\ref{sec:numerical}, we introduce the numerical framework, including the reduced perturbation system, the stochastic sources, and the Fokker--Planck formulation used to compute the scalar power spectrum.
In Sec.~\ref{sec:r_and_ns}, we compute $n_s$ and $r$ and confront them with CMB observations.
Finally, in Sec.~\ref{sec:conclusions}, we summarize our results.

\section{Two-scalar-field cosmological model\label{sec:background}}

Inflation, dark energy, and dark matter can be unified within a two-scalar-field cosmological model, in which one field drives WI at early times and subsequently behaves as dark matter, while the other field accounts for dark energy at late times \cite{Sa:2020fvn, Sa:2021eft}.

This model is described by the action
\begin{align}
 S = & \int d^4x \sqrt{-g} \bigg[
  \frac{R}{2\sigma^2} - \frac12 (\nabla \phi)^2
    \nonumber
  \\
  & - \frac12 e^{-s\sigma\phi} (\nabla \xi)^2
  - e^{-b\sigma\phi} V(\xi) \bigg],
 \label{action 2SF}
\end{align}
with the potential
\begin{equation}
 V(\xi)  = V_a + \frac12 m^2 \xi^2,
  \label{potential xi}
\end{equation}
where $g$ is the determinant of the metric $g_{\mu\nu}$, $R$ is the Ricci scalar, $\xi$ and $\phi$ are scalar fields, $s$ and $b$ are dimensionless parameters, $V_a$ and $m$ are constants of mass dimension $4$ and $1$, respectively, $\sigma$ denotes the reduced inverse Planck mass, $\sigma=\sqrt{8\pi}/m_p$, and we use natural units, $\hbar=c=k_B=1$.

According to this unification scenario, WI is driven by the scalar field $\xi$.
Soon after inflation ends, $\xi$ decouples from radiation and begins to oscillate about the minimum of its quadratic potential, thereafter behaving like a cold dark matter fluid.
Following a radiation-dominated era, this dark matter---along with standard baryonic matter---comes to dominate the evolution of the Universe.
Lastly, the second scalar field $\phi$ emerges as the dominant component, driving the present era of accelerated expansion.

This model qualitatively reproduces the successive eras of the Universe's evolution with appropriate durations \cite{Sa:2020fvn, Sa:2021eft} and, at low redshifts, agrees quantitatively with observational data, provided that the parameters $s$ and $b$ assume values within the intervals \cite{DAgostino:2021vvv}
\begin{equation}
s = 0.36^{+0.18}_{-0.26} \quad \mbox{and} \quad b= 0.01^{+0.34}_{-0.24} \,.
\label{values_s_b}
\end{equation}

In general, both the inflaton field $\xi$ and the dark-energy field $\phi$ can transfer energy to a radiation bath during WI.
This ensures that the radiation energy density remains substantial throughout the inflationary expansion, allowing a smooth transition to the radiation-dominated era without a distinct post-inflationary reheating phase.
However, a direct energy transfer from the dark-energy field $\phi$ to the radiation bath is not essential for implementing the WI scenario.
Therefore, for the sake of simplicity, we neglect any direct energy exchange between $\phi$ and radiation in this paper, assuming that the corresponding dissipation coefficient vanishes
\begin{equation}
 \Gamma_{\phi} = 0.
\end{equation}
In this case, the radiation bath is sustained exclusively by the inflaton $\xi$, with dissipation described by the coefficient 
\cite{Sa:2020fvn}
\begin{align}
 \Gamma_{\xi} = f_\xi \times \left\{
    \begin{aligned}
     & T^p,
     & T\geq T_\texttt{E},
    \\
     & T^p  \exp \left[ 1- \left( \frac{T_\texttt{E}}{T} \right)^q \right],
     & T\leq T_\texttt{E},
    \end{aligned}
 \right. 
 \label{gamma}
\end{align}
where $f_\xi$ is a constant of mass dimension $1-p$, and the parameters $p$ and $q$ (with $q>0$) determine the temperature dependence of the dissipation coefficient.
The temperature $T_\texttt{E}$ is determined dynamically by the background evolution as the temperature of the radiation bath at the moment when the energy density of $\xi$ equals that of radiation, i.e., when radiation takes over as the dominant component of the Universe.

The functional form of $\Gamma_\xi$ given by Eq.~(\ref{gamma}) is phenomenological, motivated by the Boltzmann suppression of dissipative effects once the temperature drops below the mass of the fields mediating the energy transfer from the inflaton, as found for on-shell contributions in supersymmetric models~\cite{Bastero-Gil:2012akf} and in the Warm Little Inflaton scenario~\cite{Rosa:2018iwu} (see also Ref.~\cite{Bastero-Gil:2019gao}).
For $q=1$, its low-temperature branch reduces to $\Gamma_\xi \propto T^p \exp\{-T_\texttt{E}/T\}$, with $T_\texttt{E}$ playing the role of the mediator mass, while larger values of $q$ yield a sharper switch-off of dissipation.
Specific microphysical realizations correspond to definite values of $p$: $p=-1$ for the renormalizable effective field theory of inflation~\cite{Bastero-Gil:2019gao}; $p=1$ for the Warm Little Inflaton scenario~\cite{Bastero-Gil:2016qru}; and $p=3$ for the low-temperature supersymmetric two-stage decay mechanism~\cite{Bastero-Gil:2010dgy}.

Assuming a spatially flat FLRW metric, 
\begin{align}
ds^2 =g_{\mu\nu}(t) dx^\mu dx^\nu = -dt^2 + a^2(t) \, dx_idx^i,
  \label{FRW_metric}
\end{align}
where $a(t)$ is the scale factor, the equations of motion for the scalar fields $\xi(t)$ and $\phi(t)$ and the radiation energy density $\rho_\texttt{R}(t)$ during the first stage of evolution---comprising the inflationary period and the transition to the radiation-dominated era---are given by
\begin{align}
 & \ddot{\xi} + 3 H \dot{\xi} - s\sigma \dot{\phi}\dot{\xi}
 + V^\prime e^{(s-b)\sigma\phi}
 = - \Gamma_\xi \dot{\xi} e^{s\sigma\phi}, \label{ddot-xi-back}
  \\
 & \ddot{\phi} + 3H\dot{\phi} + \frac{s\sigma}{2}
 \dot{\xi}^2 e^{-s\sigma\phi}
 -b\sigma V e^{-b \sigma \phi}
 = 0, \label{ddot-phi-back}
  \\
& \dot{\rho}_\texttt{R} + 4H\rho_\texttt{R} = \Gamma_\xi \dot{\xi}^2, \label{dot-rho-back}
\end{align}
where $\rho_\texttt{R}=(\pi^2 g_\star/30)T^4$ and $g_\star=106.75$ represents the total number of effective relativistic degrees of freedom.
The Einstein equations read
\begin{align}
H^2 &= \frac{\sigma^2}{3} \bigg(
 \frac{\dot{\phi}^2}{2} + \frac{\dot{\xi}^2}{2} e^{-s\sigma\phi}
 + V e^{-b\sigma\phi} + \rho_\texttt{R} \bigg), \label{H}
   \\
\dot{H} &=  - \frac{\sigma^2}{2} \bigg( \dot{\phi}^2 +
 \dot{\xi}^2 e^{-s\sigma\phi}  + \frac43 \rho_\texttt{R} \bigg).
 \label{dotH}
\end{align}
In the above equations, $H=\dot{a}/a$ is the Hubble parameter, an overdot denotes a derivative with respect to cosmic time $t$, while a prime represents a derivative with respect to $\xi$.
In these equations, radiation is described as a perfect fluid with an equation of state $p_\texttt{R}=\rho_\texttt{R}/3$.

The scalar fields $\xi$ and $\phi$ can also be described as perfect fluids, with energy densities and pressures
\begin{align}
 \rho_\xi &= \frac12 \dot{\xi}^2 e^{-s\sigma\phi}
 + \frac12 m^2 e^{-b\sigma\phi} \xi^2,
  \label{rho_xi_back}
\\
  p_\xi &= \frac12 \dot{\xi}^2 e^{-s\sigma\phi}
 - \frac12 m^2 e^{-b\sigma\phi} \xi^2,
    \label{p_xi_back}
\end{align}
and
\begin{align}
\rho_\phi &= \frac12 \dot{\phi}^2 + V_a e^{-b\sigma\phi},
  \label{rho_phi_back}
 \\
  p_\phi &= \frac12 \dot{\phi}^2 - V_a e^{-b\sigma\phi}.
  \label{p_phi_back}
\end{align}
In terms of these quantities, Eqs.~(\ref{ddot-xi-back}) and (\ref{ddot-phi-back}) take the form
\begin{align}
  \dot{\rho}_\xi
  + 3 H (\rho_\xi+p_\xi)
   &= - \Gamma_\xi \dot\xi^2 + W_\xi, \\
\dot{\rho}_\phi
  + 3 H (\rho_\phi+p_\phi)
   &= W_\phi,
\end{align}
where the source terms
\begin{equation}
    W_\xi = -W_\phi = \frac12 \sigma \dot{\phi} \left( s\, \dot{\xi}^2 e^{-s\sigma\phi} - b\, m^2 \xi^2  e^{-b\sigma\phi} \right)
    \label{interaction}
\end{equation}
account for the direct coupling between the scalar fields $\xi$ and $\phi$.

Following the inflationary epoch, the dissipation coefficient $\Gamma_\xi$ experiences exponential suppression (see Eq.~(\ref{gamma})) and soon afterward becomes negligible.
In the absence of dissipation, the field $\xi$ decouples from the radiation bath and undergoes rapid oscillations about the minimum of its quadratic potential, effectively behaving as a nonrelativistic pressureless dark-matter fluid \cite{Turner:1983he}.
During this second stage of evolution---comprising the \mbox{radiation-,} matter-, and dark-energy-dominated eras---the energy densities of radiation and the scalar field $\xi$ evolve as
\begin{equation}
\rho_\texttt{R}=\rho_\texttt{R0} \left( \frac{a_0}{a} \right)^4
\end{equation}
and \cite{Sa:2020fvn}
\begin{equation}
\rho_\xi=\rho_\texttt{DM0} \left(  \frac{a_0}{a} \right)^3 e^{\frac{(s-b)\sigma}{2}(\phi-\phi_0)},
    \label{xi 2}
\end{equation}
where the subscript $0$ indicates present-time values.

When standard baryonic matter, described as a perfect fluid with pressure $p_\texttt{BM}=0$ and energy density $\rho_\texttt{BM} = \rho_\texttt{BM0} (a_0/a)^3$,  is also included, the equation for the dark-energy field becomes
\begin{align}
&\ddot{\phi}+3H\dot\phi-b\sigma V_a e^{-b\sigma\phi}  \nonumber
\\
& + \frac12 \sigma (s-b) \rho_\texttt{DM0} \left( \frac{a_0}{a} \right)^3 e^{\frac{(s-b)\sigma}{2}(\phi-\phi_0)} = 0\,,
    \label{ddot-phi-2}
\end{align}
while the Einstein equations are now given by
\begin{align}
  H^2=&\frac{\sigma^2}{3} \bigg[ \frac{\dot{\phi}^2}{2} + V_a e^{-b\sigma\phi} + \rho_\texttt{R0} \left( \frac{a_0}{a} \right)^4 \nonumber
\\
& + \left( \rho_\texttt{BM0} + \rho_\texttt{DM0} e^{\frac{(s-b)\sigma}{2}(\phi-\phi_0)} \right) \left( \frac{a_0}{a} \right)^3\bigg],
    \label{H-2}
    \\
 \dot{H}=& -\frac{\sigma^2}{2} \bigg[ \dot{\phi}^2 +
 \left( \rho_\texttt{BM0} + \rho_\texttt{DM0} e^{\frac{(s-b)\sigma}{2}(\phi-\phi_0)} \right) \left( \frac{a_0}{a} \right)^3 \nonumber
    \\ &+ \frac43 \rho_\texttt{R0} \left( \frac{a_0}{a} \right)^4 \bigg].
    \label{dotH-2}
\end{align}
Agreement with current cosmological measurements \cite{Fixsen:2009ug, Planck:2018vyg} requires $\rho_\texttt{R0}=9.02\times 10^{-128}\,m_p^4$, $\rho_\texttt{BM0}=8.19\times10^{-125}\,m_p^4$, $\rho_\texttt{DM0}=4.25\times10^{-124}\,m_p^4$, and
$\rho_\texttt{DE0}\equiv\frac12\dot{\phi}_0^2+V_a e^{-b\sigma\phi_0}=1.13\times10^{-123}\,m_p^4$, corresponding to a Hubble constant of $H_0=67\,\mbox{km}\,\mbox{s}^{-1}\mbox{Mpc}^{-1}$.
These values satisfy $\rho_\texttt{R0}+\rho_\texttt{BM0}+\rho_\texttt{DM0}+\rho_\texttt{DE0}=\rho_{c0}$, where $\rho_{c0}\equiv 3H_0^2/\sigma^2$ is the present critical energy density, as required by the spatial flatness assumed in Eq.~\eqref{FRW_metric}.

During the second stage of evolution, the source terms in Eq.~(\ref{interaction}) simplify to \cite{Sa:2021eft}
\begin{equation}
W_\xi = -W_\phi = \frac12\sigma(s-b)\dot\phi\rho_\xi.
\label{interaction-late-time}
\end{equation}
Cosmological models featuring this type of interaction term---proportional to $\dot{\phi}\rho_\texttt{DM}$, where $\phi$ is a quintessence field with an exponential potential and $\rho_\texttt{DM}$ is the dark-matter energy density---have been extensively investigated in the literature (see, for example, Refs.~\cite{Amendola:1999er, Gumjudpai:2005, Boehmer:2009roi, Amendola:2014kwa, Bahamonde:2017ize, Potting:2022}).

The evolution equations for the two stages, Eqs.~(\ref{ddot-xi-back})--(\ref{dotH}) and (\ref{ddot-phi-2})--(\ref{dotH-2}), are integrated numerically; for numerical convenience, we replace the cosmic time $t$ with the variable $u=\ln(a/a_0)$, where $a_0\equiv a(u=0)$ is the present-day scale factor, and denote derivatives with respect to $u$ by a subscript $u$.
Further details on the numerical integration can be found in Ref.~\cite{Sa:2020fvn}.

As an illustrative example of the background dynamics in the two-scalar-field cosmological model, we consider a scenario defined by the following parameter values and initial conditions: $s=0.36$ and $b=0.01$, corresponding to the best-fit values in Eq.~\eqref{values_s_b} \cite{DAgostino:2021vvv}, $V_a=1.10 \times 10^{-123} \, m_p^4$, $m=8.54 \times 10^{-8} \, m_p$, $f_\xi = 4.76\times 10^{-11}\,m_p^2$, $p=-1$, $q=2$, $\xi(u_i)=1.12 \, m_p$, $\phi(u_i) = 1.99\times 10^{-4}\, m_p$, $\xi_u(u_i)=-1.91\times 10^{-2}\, m_p$, $\phi_u(u_i)=1.99\times 10^{-6}\, m_p$, and $\rho_\texttt{R}(u_i)=2.85\times
10^{-17} \,m_p^4$.

\begin{figure}[t]
 \includegraphics[scale=1.3]{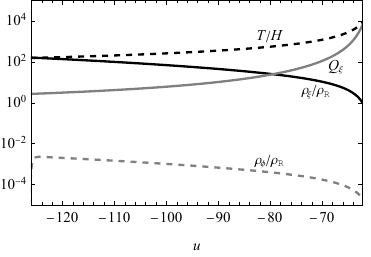}
\caption{Evolution of $Q_\xi$, $T/H$, $\rho_\xi/\rho_\texttt{R}$, and $\rho_\phi/\rho_\texttt{R}$ during the inflationary period, which extends from $u_i \simeq -126.2$ to $u_e \simeq -62.2$ ($64$ $e$-folds of expansion). 
The transition from inflation to the radiation-dominated era occurs when \mbox{$\rho_\xi/\rho_\texttt{R}=1$}.
Throughout inflation, $\rho_\xi$ exceeds $\rho_\phi$ by several orders of magnitude, $T > H$, and the system remains in the strong dissipative regime $Q_\xi > 1$.}
\label{fig1}
\end{figure}

In Fig.~\ref{fig1}, we present the evolution of the dissipation ratio
\begin{equation}
Q_\xi = \frac{\Gamma_\xi}{3H} e^{s\sigma\phi},
\label{diss-ratio}
\end{equation}
alongside the ratios $T/H$, $\rho_\xi/\rho_\texttt{R}$, and $\rho_\phi/\rho_\texttt{R}$, throughout the inflationary period.
Dissipation remains strong ($Q_\xi>1$) for the entire duration of inflation.

\begin{figure}[t]
 \includegraphics[scale=1.3]{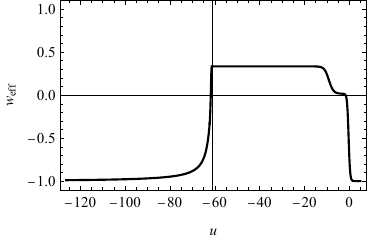}
\caption{Evolution of $w_\text{eff}$ from the beginning of the inflationary period to the present. 
The succession of evolutionary epochs---inflation, radiation dominance, matter dominance, and dark-energy dominance---is clearly visible.
The radiation-dominated era encompasses Big Bang nucleosynthesis, and the subsequent era dominated by dark and baryonic matter lasts long enough to allow the formation of large-scale structure.}
\label{fig2}
\end{figure}

Figure~\ref{fig2} shows the evolution of the effective equation-of-state parameter.
This parameter is defined as $w_\text{eff} = \sum p_x / \sum \rho_x$, where $x$ labels the cosmic components across different evolutionary stages.
During the first stage, spanning the inflationary period and the transition to radiation dominance, $x$ includes the inflaton $\xi$, the dark-energy field $\phi$, and radiation; during the second stage, extending from the onset of radiation dominance to the present day, $x$ comprises cold dark matter (field $\xi$ oscillating rapidly about the minimum of its potential), the dark-energy field $\phi$, standard baryonic matter, and radiation.

The background evolution shown in Figs.~\ref{fig1} and \ref{fig2} corresponds to the scenario labeled $Q_{\xi\ast}=3.23$ in Fig.~\ref{fig:rXns}, where $Q_{\xi\ast}$ is the dissipation ratio at Hubble crossing of the pivot scale (see Sec.~\ref{sec:r_and_ns}).
Once scalar perturbations are taken into account, this scenario yields values of $n_s$ and $r$ lying well inside the 68\% C.L.\ region of the $(n_s, r)$ plane.

\section{Equations for scalar perturbations\label{sec:perturbation}}

In this section, we derive the linear scalar perturbation equations for the two-scalar-field WI model introduced in Sec.~\ref{sec:background}.
We treat the inflaton, radiation fluid, and metric perturbations as a coupled system while neglecting the dark-energy field perturbations.
The dissipative energy transfer and the associated stochastic sources are included consistently in our prescription.
Finally, we derive the linearized Einstein equations, which close the system.
The resulting gauge-ready formulation provides the framework for the numerical methods developed in Sec.~\ref{sec:numerical}.

\subsection{Scalar-perturbed FLRW metric}

In WI, primordial density perturbations are sourced predominantly by thermal fluctuations of the radiation bath, which the dissipative coupling transfers to the inflaton, rather than by the quantum vacuum fluctuations that dominate in CI. 
Their computation therefore requires following the perturbations of the radiation fluid alongside those of the inflaton, and supplementing the inflaton perturbation equation with stochastic thermal and quantum noise terms.
Here, stochastic contributions associated with the dissipative energy transfer to the radiation bath are also introduced.

In standard, single-field WI, the perturbed equations are well known (see, for example, the review article \cite{Kamali:2023lzq} and Refs.~\cite{Graham:2009bf, Bastero-Gil:2011rva}).

Here, we generalize these equations to the case of the two-scalar-field cosmological model described in Sec.~\ref{sec:background}, in which the inflaton is directly coupled to a second scalar field that accounts for dark energy at later stages of the Universe's evolution\footnote{For general gauge-ready formulations of scalar perturbations in cosmologies containing multiple interacting fluids and scalar fields, see Refs.~\cite{Hwang:2001fb, Malik:2002jb}.}.

To this end, we consider the most general scalar-perturbed FLRW metric, given by the line element \cite{Kamali:2023lzq}
\begin{align}
ds^2 &= \bar{g}_{\mu\nu}dx^\mu dx^\nu =-(1+2\alpha)\,dt^2
- 2a\,\partial_i \beta\, dx^i dt \nonumber
    \\
&+ a^2\Big[\delta_{ij}(1+2\varphi) + 2\partial_i\partial_j \gamma \Big] dx^i dx^j,    \label{perturbed_metric}
\end{align}
where the scale factor $a$ is a function of the cosmic time $t$, while the metric perturbations $\alpha$, $\beta$, $\varphi$, and $\gamma$ are functions of $t$ and $x^i$.

The metric tensor $\bar{g}_{\mu\nu}(t,x^i)$ can be decomposed into a homogeneous background and a perturbation,
\begin{equation}
   \bar{g}_{\mu\nu}(t,x^i) = g_{\mu\nu}(t) + \delta g_{\mu\nu}(t,x^i),
\label{decomposition}
\end{equation}
where $g_{\mu\nu}$ is given by Eq.~(\ref{FRW_metric}) and the contravariant perturbation $\delta g^{\mu\nu}$ is related to its covariant counterpart to linear order via
\begin{equation}
   \delta g^{\mu\nu}=- g^{\mu\alpha} g^{\nu\beta} \delta g_{\alpha\beta}.
\end{equation}

Whereas in Sec.~\ref{sec:background} all quantities referred to the background, here every physical quantity is decomposed, like the metric, into a background part and a perturbation,
\begin{equation}
\bar{X}(t,x^i) = X(t) + \delta X(t,x^i),
\end{equation}
where $X$ denotes a generic quantity, whether it be a scalar, vector, or tensor.
This notation is reserved for quantities with a background counterpart; quantities defined only at the perturbative level, such as the metric perturbations $\alpha$, $\beta$, $\varphi$, and $\gamma$, as well as the momentum potentials and the comoving curvature perturbation introduced below (see Eqs.~\eqref{def_psi_1u}--\eqref{def_psi_2v} and \eqref{R_total_momentum}, respectively), are written without $\delta$, following standard usage.

\subsection{Conservation laws}

The scalar-field sector of the action (\ref{action 2SF}) is described by the Lagrangian density
\begin{equation}
 {\cal L} =- \frac12 \bar{g}^{\mu\nu}\partial_\mu\bar{\phi}\partial_\nu\bar{\phi}- \frac12 e^{-s\sigma\bar{\phi}} \bar{g}^{\mu\nu}\partial_\mu\bar{\xi}\partial_\nu\bar{\xi}
  - e^{-b\sigma\bar{\phi}} V(\bar{\xi}).
 \label{lagrangian}
\end{equation}

The corresponding energy-momentum tensor is given by
\begin{align}
\bar{T}^\nu{\!}_\mu &= \bar{g}^{\nu\lambda}\partial_\lambda\bar{\phi} \partial_\mu\bar{\phi} 
+ e^{-s\sigma\bar{\phi}} \bar{g}^{\nu\lambda} 
\partial_\lambda\bar{\xi} \partial_\mu\bar{\xi} 
-\delta_\mu^\nu
\bigg( \frac12 \bar{g}^{\alpha\beta} \partial_\alpha\bar{\phi} \partial_\beta\bar{\phi} 
\nonumber
\\
& \hspace{10mm}
 + \frac12 e^{-s\sigma\bar{\phi}} \bar{g}^{\alpha\beta} \partial_\alpha\bar{\xi} \partial_\beta\bar{\xi}
+ e^{-b\sigma\bar{\phi}} V(\bar{\xi}) \bigg),
\end{align}
from which we define the energy density and pressure of the scalar fields as
\begin{equation}
    \bar{\rho}_{\text{sf}} = \bar{\rho}_\xi + \bar{\rho}_\phi = -\bar{T}^0{\!}_0 \quad \mbox{and} \quad \bar{p}_{\text{sf}} = \bar{p}_\xi+ \bar{p}_\phi = \frac{\bar{T}^i{\!}_i}{3}.
\end{equation}

These definitions yield, at the background level, Eqs.~(\ref{rho_xi_back})--(\ref{p_phi_back}), while at the perturbation level, they result in the expressions
\begin{align}
\delta\rho_{\xi} &= \left( \dot{\xi} \dot{\delta\xi} -\alpha \dot{\xi}^2 \right) e^{-s\sigma\phi} + V^\prime \delta\xi e^{-b\sigma\phi}, \label{rho_xi_pert}
\\
\delta p_{\xi} &= \left( \dot{\xi} \dot{\delta\xi} -\alpha \dot{\xi}^2 \right) e^{-s\sigma\phi} - V^\prime \delta\xi e^{-b\sigma\phi}, \label{p_xi_pert}
\\
\delta\rho_{\phi} &= -\alpha \dot{\phi}^2, \label{rho_phi_pert}
\\
\delta p_{\phi} &= -\alpha \dot{\phi}^2. \label{p_phi_pert}
\end{align}

In Eqs.~(\ref{rho_xi_pert})--(\ref{p_phi_pert}), we have set $\delta\phi=0$, an additional simplifying assumption adopted on top of $\bar\Gamma_\phi=0$, already introduced in Sec.~\ref{sec:background}.
Because of the direct coupling between the two fields, a perturbation $\delta\xi$ will, in general, source a perturbation $\delta\phi$; here, however, we neglect $\delta\phi$ and hence its backreaction on $\delta\xi$ and on the metric perturbations. 
Consistently, $\delta\phi$ is also dropped from the remaining perturbation equations. 
With $\delta\phi=0$,  Eq.~(\ref{phi-perturb}) below is no longer an evolution equation but a relation among $\alpha$, $\beta$, $\varphi$, $\gamma$, and $\delta\xi$; in Sec.~\ref{subsec-consistency-check}, we evaluate its residual along the numerical solution to check \textit{a posteriori} that this approximation is consistent. 
The more general scenario, with $\bar\Gamma_\phi\neq0$ and $\delta\phi$ fully taken into account, is left for future work.

The expressions for the background and perturbed energy densities and pressures can also be obtained by treating the scalar fields $\bar{\xi}$ and $\bar{\phi}$ as perfect fluids with energy-momentum tensors given by \cite{Faraoni:2012hn, Harko:2021gnz}
\begin{align}
    \bar{T}_\xi^{\mu\nu} &= (\bar{\rho}_\xi+\bar{p}_\xi) \bar{u}^\mu \bar{u}^\nu + \bar{p}_\xi \bar{g}^{\mu\nu}, \\
    \bar{T}_\phi^{\mu\nu} &= (\bar{\rho}_\phi +\bar{p}_\phi) \bar{v}^\mu \bar{v}^\nu + \bar{p}_\phi \bar{g}^{\mu\nu},
\end{align}
with timelike four-velocities
\begin{align}
    \bar{u}^\alpha &=- \epsilon_\xi\, 
    \frac{   \bar{g}^{\alpha\beta} \partial_\beta     \bar{\xi} e^{- s\sigma\bar{\phi}/2}}{ \sqrt{\bar{\rho}_\xi + \bar{p}_\xi}}, \label{4vel_u}
    \\
    \bar{v}^\alpha &=-   \epsilon_\phi\, \frac{   \bar{g}^{\alpha\beta}\partial_\beta     \bar{\phi}}{\sqrt{\bar{\rho}_\phi + \bar{p}_\phi}}, \label{4vel_v}
\end{align}
where $\epsilon_\xi \equiv \sgn(\dot\xi)$ and $\epsilon_\phi \equiv \sgn(\dot\phi)$ ensure that $u^0 = v^0 = 1$, irrespective of the direction in which the fields roll.

In what follows, we assume that the scalar field $\xi$ and the radiation fluid share the same four-velocity in the background state, $u^\alpha$, and in the perturbed zeroth (temporal) component, $\delta u^0$, while exhibiting distinct perturbed spatial components, which we denote by $\delta u^i$ and $\delta w^i$, respectively.

The conditions $\bar{u}^\alpha\bar{u}_\alpha=-1$ and $\bar{v}^\alpha\bar{v}_\alpha=-1$ fix the zeroth components of the perturbed four-velocities to be $\delta u^0=\delta u_0=\delta v^0=\delta v_0=-\alpha$, but impose no restrictions on the spatial components $\delta u^i$, $\delta w^i$, and $\delta v^i$.
We choose to express them in terms of new functions, $\psi_\xi$, $\psi_\texttt{R}$, and $\psi_\phi$, respectively.
This parametrization is analogous to the introduction of a potential, since in the scalar sector the corresponding momentum perturbations are expressed as spatial gradients of scalar quantities.
More specifically, $\psi_\xi$, $\psi_\texttt{R}$, and $\psi_\phi$ play the role of momentum potentials and are introduced as
\begin{align}
    (\rho_\xi + p_\xi) \delta u^i &=(\rho_\xi + p_\xi)\frac{\partial^i\beta}{a} + \frac{\partial^i\psi_\xi}{a^2}, \label{def_psi_1u} \\
    (\rho_\texttt{R} + p_\texttt{R}) \delta w^i &=(\rho_\texttt{R} + p_\texttt{R})\frac{\partial^i\beta}{a} + \frac{\partial^i\psi_\texttt{R}}{a^2}, \label{def_psiR_1u} \\
    (\rho_\phi + p_\phi)\delta v^i &=(\rho_\phi + p_\phi)\frac{\partial^i\beta}{a} + \frac{\partial^i\psi_\phi}{a^2}, \label{def_psi_1v}
\end{align}
or, equivalently, through
\begin{align}
(\rho_\xi + p_\xi) \delta u_i &= \partial_i \psi_\xi,  \label{def_psi_2u} \\
(\rho_\texttt{R} + p_\texttt{R}) \delta w_i &= \partial_i \psi_\texttt{R},  \label{def_psiR_2u} \\
(\rho_\phi + p_\phi) \delta v_i &= \partial_i \psi_\phi. \label{def_psi_2v}
\end{align}

The dissipative source terms are defined as
\begin{equation}\label{eq:diss_source}
    \bar{S}_\texttt{R}^\nu=-\bar{S}^\nu_\xi=-\bar{g}^{\mu\nu}\bar{\Gamma}_\xi \bar{u}^\alpha\partial_\alpha\bar{\xi}\partial_\mu\bar{\xi},
\end{equation}
yielding, for the background,
\begin{align}
S_\texttt{R}^0=-S_\xi^0 & = \Gamma_\xi \dot{\xi}^2, 
\\
S_\texttt{R}^i=-S_\xi^i & = 0,
\end{align}
in agreement with Eqs.~(\ref{ddot-xi-back}) and (\ref{dot-rho-back}), while the temporal and spatial components of the perturbations are
\begin{align}
\delta S_\texttt{R}^0=-\delta S_\xi^0 &= \delta \Gamma_\xi \dot{\xi}^2 + 2\Gamma_\xi \dot{\xi} \dot{\delta\xi} - 3\alpha\Gamma_\xi \dot{\xi}^2,
\\
\delta S_\texttt{R}^i=-\delta S_\xi^i &= \frac{\Gamma_\xi}{a}\dot{\xi}^2 \partial^i\beta - \frac{\Gamma_\xi}{a^2}\dot{\xi}\partial^i\delta\xi.
\end{align}
Since we assume there is no direct energy exchange between the scalar field $\phi$ and radiation, the corresponding source term $\bar{S}^\nu_\phi$ is identically zero.

The conservation law $\bar{\nabla}_\mu \bar{T}^{\mu\nu}=0$ can be written as
\begin{align}
\bar{\nabla}_\mu\bar{T}_\xi^{\mu\nu} &= \bar{S}_\xi^\nu+\bar{W}_\xi^\nu, \label{cl1} \\
\bar{\nabla}_\mu\bar{T}_\phi^{\mu\nu} &= \bar{W}_\phi^\nu, \label{cl2} \\
\bar{\nabla}_\mu\bar{T}_\texttt{R}^{\mu\nu} &= \bar{S}_\texttt{R}^\nu,
\label{cl3}
\end{align}
where, besides the dissipative terms $\bar{S}^\nu_\xi$ and $\bar{S}_\texttt{R}^\nu$, there are also the source terms  
\begin{align}
\bar{W}_\xi^\nu = -\bar{W}_\phi^\nu =& \frac12 \sigma \bar{g}^{\mu\nu} \partial_\mu \bar{\phi} \Big(s\, e^{-s\sigma\bar{\phi}} \bar{g}^{\alpha\beta} \partial_\alpha\bar{\xi} \partial_\beta\bar{\xi} \nonumber
\\
&+ b\, m^2 \bar{\xi}^2 e^{-b\sigma\bar{\phi}} \Big),
\end{align}
which describe the direct coupling between the scalar fields $\bar{\xi}$ and $\bar\phi$.
At the background level, the zeroth components of $W_\xi^\nu$ and $W_\phi^\nu$ are given by Eq.~(\ref{interaction}) and, after the inflationary period, when the scalar field $\xi$ oscillates rapidly about the minimum of its quadratic potential, by Eq.~(\ref{interaction-late-time}).
The zeroth components of the perturbed source terms, $\delta W_\xi^\nu$ and $\delta W_\phi^\nu$, are
\begin{align}
    \delta W_\xi^0 = -\delta W_\phi^0 &= \sigma\dot\phi
\Big[ s\, \dot\xi \left( -2\alpha\dot{\xi} + \dot{\delta\xi} \right) e^{-s\sigma\phi} \nonumber 
\\
& + b\, m^2 \xi \left( \alpha \xi- \delta \xi \right) e^{-b\sigma\phi} \Big].
\end{align}

Conservation laws (\ref{cl1})--(\ref{cl3}) yield, for $\nu=0$, the background equations (\ref{ddot-xi-back})--(\ref{dot-rho-back}) and the perturbed equations
\begin{widetext}
\begin{align}
    \dot{\delta\rho_\xi} + 3H(\delta\rho_\xi + \delta p_\xi) &= -\left(\rho_\xi+p_\xi\right) \left( 3\dot\varphi + \partial_i\partial^i\dot\gamma + \frac{\partial_i\partial^i\beta}{a} \right) -\frac{\partial_i\partial^i\psi_\xi}{a^2} +\delta S^0_\xi+2\alpha S^0_\xi + \delta W^0_\xi + 2\alpha W^0_\xi,
    \label{rho_perturbed_xi}  \\
    \dot{\delta\rho_\phi} + 3H(\delta\rho_\phi + \delta p_\phi) &= -\left(\rho_\phi+p_\phi\right) \left( 3\dot\varphi + \partial_i\partial^i\dot\gamma + \frac{\partial_i\partial^i\beta}{a} \right) -\frac{\partial_i\partial^i\psi_\phi}{a^2}  +\delta W^0_\phi + 2\alpha W^0_\phi,
    \label{rho_perturbed_phi}  \\
    \dot{\delta\rho_\texttt{R}} + 3H(\delta\rho_\texttt{R} + \delta p_\texttt{R}) &= -\left(\rho_\texttt{R} + p_\texttt{R}\right) \left( 3\dot\varphi + \partial_i\partial^i\dot\gamma + \frac{\partial_i\partial^i\beta}{a} \right) -\frac{\partial_i\partial^i\psi_\texttt{R}}{a^2} +\delta S^0_\texttt{R} + 2\alpha S^0_\texttt{R}.
    \label{rho_perturbed_r}
\end{align}
\end{widetext}

To express $\psi_\xi$ in terms of $\delta\xi$, we perturb Eq.~(\ref{4vel_u}), obtaining
\begin{align}
\sqrt{\rho_\xi+p_\xi}\,\delta u^\alpha =&  \epsilon_\xi\,   e^{-s\sigma\phi/2}\Big[ -\partial_\beta\delta\xi\, g^{\beta\alpha}
-\partial_\beta\xi\,\delta g^{\beta\alpha}
\nonumber
\\
&+\partial_\beta\xi\, g^{\beta\alpha}\,
\frac{\delta\rho_\xi+\delta p_\xi}{2(\rho_\xi+p_\xi)}\Big],
\end{align}
which, using Eqs.~(\ref{rho_xi_back}), (\ref{p_xi_back}), (\ref{rho_xi_pert}), and (\ref{p_xi_pert}), simplifies to
\begin{equation}
    \dot{\xi}^2 \delta u^\alpha = -g^{\beta\alpha} \dot{\xi} \partial _\beta \delta\xi - \delta g^{\beta\alpha} \dot\xi \partial _\beta\xi + g^{\beta\alpha} \partial_\beta \xi (\dot{\delta\xi}-\alpha\dot{\xi}).
\end{equation}
Inserting the spatial component 
\begin{equation}
    \dot{\xi}^2 \delta u^i=-\frac{\dot\xi}{a^2}\partial^i \delta\xi + \frac{\dot\xi^2}{a}  \partial^i\beta
\end{equation}
into Eq.~(\ref{def_psi_1u}) yields
\begin{equation}
    \partial^i\psi_\xi= - e^{-s\sigma\phi} \dot\xi \partial^i \delta \xi.
    \label{psi_xi}
\end{equation}

Similarly, for $\psi_\phi$, we obtain
\begin{equation}
    \partial^i\psi_\phi=- \dot\phi \partial^i \delta \phi =0.
    \label{psi_phi}
\end{equation}

Using these results, Eqs.~(\ref{rho_perturbed_xi})--(\ref{rho_perturbed_r}) can be written as
\begin{align}
   & \ddot{\delta\xi}+3H\dot{\delta \xi} + \left( V^{\prime\prime} e^{(s-b)\sigma\phi} - \frac{\partial_i\partial^i}{a^2} \right) \delta\xi = -\Gamma_\xi \dot{\delta\xi}e^{s\sigma\phi} \nonumber
\\
& \hspace{10mm} + 2\alpha\ddot{\xi} + \dot\xi \left( \dot\alpha + 3H\alpha + \kappa - \delta \Gamma_\xi e^{s\sigma\phi}  + \alpha \Gamma_\xi e^{s\sigma\phi} \right) \nonumber
\\
& \hspace{10mm} - s \sigma \dot\phi \left( 2\alpha \dot\xi-\dot{\delta\xi}  \right),
    \label{xi-perturb}
\\
&    \dot\phi\left( \dot\alpha  -3H\alpha + \kappa \right) = s \sigma \dot\xi \dot{\delta\xi} e^{-s\sigma\phi} - b\, \sigma m^2 \xi \delta\xi e^{-b\sigma\phi} \nonumber
\\
& \hspace{10mm}
- 2 b\, \sigma V(\xi) \alpha e^{-b\sigma\phi}, \label{phi-perturb}
\\
&    \dot{\delta\rho_\texttt{R}} + 4H\delta\rho_\texttt{R} = - \frac43\rho_\texttt{R} \left( 3H\alpha - \kappa \right) -\frac{\partial_i\partial^i\psi_\texttt{R}}{a^2} \nonumber
\\
& \hspace{10mm}  + \delta\Gamma_\xi \dot\xi^2 + 2\Gamma_\xi\dot\xi\dot{\delta\xi} - \alpha \Gamma_\xi\dot\xi^2.
    \label{rad-perturb}
\end{align}

In the above equations, we have introduced the notation 
\begin{align}
    \chi&\equiv a(\beta+a\dot\gamma),     \label{shear scalar}
\\
    \kappa&\equiv3(-\dot\varphi+H\alpha)-\frac{1}{a^2}\partial^i\partial_i\chi,
    \label{def-kappa}
\end{align}
where $\chi$ denotes the scalar shear variable and $\kappa$ the perturbation of the trace of the extrinsic curvature.

The above set of equations must be supplemented with an equation for the time evolution of $\psi_\texttt{R}$, which we obtain from the spatial component of the conservation law (\ref{cl3}), namely, 
\begin{equation}
\dot{\psi}_\texttt{R}+3H\psi_\texttt{R}+\frac{\delta\rho_\texttt{R}}{3}+\frac43 \rho_\texttt{R} \alpha =-\Gamma_\xi \dot\xi\delta\xi.
   \label{psirad-perturb}
\end{equation}

The perturbed evolution equations (\ref{xi-perturb}), (\ref{rad-perturb}), and (\ref{psirad-perturb}) reduce to the corresponding equations for standard WI (Eqs.~(63), (60), and (61) of Ref.~\cite{Kamali:2023lzq}, respectively) in the limit $\phi\to0$.

We now assume that $\delta\xi$ is also driven by stochastic quantum and thermal white-noise terms, $\zeta_Q$ and $\zeta_T$, respectively, in direct analogy with the treatment of the standard single-field WI developed in Ref.~\cite{Ramos:2013nsa} and reviewed in Ref.~\cite{Kamali:2023lzq}.
This extension is justified by inspection of Eq.~(\ref{xi-perturb}); if we collect the terms proportional to $\dot{\delta\xi}$ and invoke the condition $|s\sigma\dot\phi| \ll 3H,\,\Gamma_\xi e^{s\sigma\phi}$---equivalent to the requirement that $\phi$ vary slowly compared to the dissipative and Hubble timescales governing $\xi$---the friction coefficient acting on $\delta\xi$ reduces to $3H(1+Q_\xi)$, with $Q_\xi$  defined in Eq.~(\ref{diss-ratio}).
This is structurally identical to the friction coefficient $3H(1+Q)$, with $Q=\Upsilon/(3H)$, appearing in the inflaton perturbation equation of Refs.~\cite{Ramos:2013nsa,Kamali:2023lzq}, with $\Gamma_\xi e^{s\sigma\phi}$ playing the role of $\Upsilon$.
Since the derivation of the noise correlation functions in Refs.~\cite{Ramos:2013nsa, Kamali:2023lzq} depends on the perturbation equation only through this friction coefficient---fixing, via the Green's-function solution built from Bessel functions of order $\nu=3(1+Q)/2$, both the Markovian fluctuation-dissipation amplitude of the thermal noise and the enhancement factor $\sqrt{9+12\pi Q}$ appearing in the quantum-noise equation---the entire derivation carries over to the present two-scalar-field cosmological model, provided that $Q$ is everywhere replaced by $Q_\xi$ and that an additional factor $e^{s\sigma\phi}$ is introduced because of the noncanonical normalization of the inflaton fluctuation.

The dissipative interaction also induces stochastic fluctuations around the mean energy transfer $\Gamma_\xi \dot{\xi}^{2}$ to the radiation bath. 
Thus, we introduce a radiation-noise source $\zeta_\texttt{R}$.
Within our prescription, this contribution is included in the radiation energy-density equation, while no independent stochastic source is introduced in the radiation momentum equation.
We emphasize that this formulation is not univocal and it is still debated \cite{Rodrigues:2025neh}.

The correlation functions adopted below thus generalize the results of Refs.~\cite{Ramos:2013nsa, Kamali:2023lzq, Rodrigues:2025neh} to our two-scalar-field cosmological model, accounting for the direct coupling between the inflaton and dark-energy fields.

The noise terms are stochastic Gaussian sources with zero mean, $\langle\zeta_Q\rangle = \langle\zeta_T\rangle = \langle\zeta_\texttt{R}\rangle = 0$, and two-point correlation functions given by
\begin{align}
   \langle \zeta_T ({\bf k},t) \zeta_T ({\bf k^\prime},t^\prime) \rangle =& \frac{6 H Q_\xi  T e^{s\sigma\phi}}{a^3} \delta(t-t^\prime)(2\pi)^3\delta({\bf k}+{\bf k^\prime}),
    \label{eq:zetaT}
\\
   \langle \zeta_Q ({\bf k},t) \zeta_Q ({\bf k^\prime},t^\prime)\rangle =& \frac{H^2\sqrt{9+12\pi Q_\xi} \,(1+2n)e^{s\sigma\phi}}{\pi a^3} \nonumber
\\  
& \times \delta(t-t^\prime)(2\pi)^3 \delta({\bf k}+{\bf k^\prime}),
    \label{eq:zetaQ} 
\\ 
\langle \zeta_\texttt{R} ({\bf k},t) \zeta_\texttt{R} ({\bf k^\prime},t^\prime) \rangle =&  \frac{6 H Q_\xi  T \dot\xi^2 e^{-s \sigma \phi}}{a^3} \nonumber
\\
& \times \delta(t-t^\prime)(2\pi)^3\delta({\bf k}+{\bf k^\prime}),
    \label{eq:zetaR}
\end{align}
where $n$ is the occupation number of the inflaton fluctuations, which, assuming that they are in thermal equilibrium with the radiation bath (see Ref.~\cite{Kamali:2023lzq} and references therein), is given by the Bose--Einstein distribution, $n = [\exp(H/T) - 1]^{-1}$.

Including the noise terms, the inflaton and radiation fluctuation dynamics become
\begin{align}
   & \ddot{\delta\xi}+3H\dot{\delta \xi} + \left( V^{\prime\prime} e^{(s-b)\sigma\phi} - \frac{\partial_i\partial^i}{a^2} \right) \delta\xi =  \zeta_Q+\zeta_T \nonumber
\\
& \hspace{5mm} -\Gamma_\xi \dot{\delta\xi}e^{s\sigma\phi} + 2\alpha\ddot{\xi} + \dot\xi \Big( \dot\alpha + 3H\alpha + \kappa - \delta \Gamma_\xi e^{s\sigma\phi}   \nonumber
\\
& \hspace{5mm} + \alpha \Gamma_\xi e^{s\sigma\phi} \Big) - s \sigma \dot\phi \left( 2\alpha \dot\xi-\dot{\delta\xi}  \right),
    \label{xi-perturb_2}
\\
& \dot{\delta\rho_\texttt{R}} + 4H\delta\rho_\texttt{R} =\zeta_\texttt{R} - \frac43\rho_\texttt{R} \left( 3H\alpha - \kappa \right) -\frac{\partial_i\partial^i\psi_\texttt{R}}{a^2} \nonumber
\\
& \hspace{5mm}  + \delta\Gamma_\xi \dot\xi^2 + 2\Gamma_\xi\dot\xi\dot{\delta\xi} - \alpha \Gamma_\xi\dot\xi^2, \label{rad-perturb_2}
\end{align}   
respectively.

\subsection{Perturbed Einstein equations}

The perturbed Einstein equations are given by
\begin{equation}
    \delta G^\mu{\!}_\nu = \sigma^2 \delta T^\mu{\!}_\nu,
\end{equation}
where $\delta G^\mu{\!}_\nu$ is the perturbed Einstein tensor and the perturbed energy-momentum tensor $\delta T^\mu{\!}_\nu$ includes contributions from radiation and the two scalar fields $\xi$ and $\phi$.

The $(0,0)$ equation, $\delta G^0{\!}_0 = \sigma^2 \delta T^0{\!}_0$, representing the energy constraint, yields 
\begin{align}
\frac{1}{a^2} \partial_i\partial^i\varphi + H\kappa =& -\frac{\sigma^2}{2} \Big[\delta \rho_\texttt{R} + \left( \dot{\xi} \dot{\delta\xi} - \alpha \dot{\xi}^2 \right) e^{-s\sigma\phi} \nonumber
\\
& + V^\prime \delta \xi e^{-b\sigma\phi} - \alpha \dot{\phi}^2 \Big], \label{Pert_Einstein00}
\end{align}
where we have taken into account that $\delta T^0{\!}_0=-(\delta \rho_\texttt{R}+\delta \rho_\xi+\delta \rho_\phi)$ and used Eqs.~(\ref{rho_xi_pert}) and (\ref{rho_phi_pert}).

The $(0,i)$ equation, $\delta G^0{\!}_i = \sigma^2 \delta T^0{\!}_i$, providing the momentum constraint, yields
\begin{equation}
\partial_i\left( \kappa + \frac{1}{a^2} \partial_j\partial^j\chi \right) = -\frac32 \sigma^2 \delta T^0{\!}_i.
\end{equation}
Using
\begin{equation}
\delta T^0{\!}_i = (p_\texttt{R}+\rho_\texttt{R})\delta w_i+(p_\xi+\rho_\xi)\delta u_i+(p_\phi+\rho_\phi)\delta v_i
\end{equation}
together with Eqs.~(\ref{def_psi_2u})--(\ref{def_psi_2v}), (\ref{psi_xi}), and (\ref{psi_phi}), we can write the momentum constraint as
\begin{equation}
\kappa + \frac{1}{a^2} \partial_i\partial^i\chi  = \frac32 \sigma^2 \left( \dot{\xi} \delta\xi e^{-s\sigma\phi} - \psi_\texttt{R} \right).
\label{Pert_Einstein0i}
\end{equation}

From $\delta G^i{\!}_i - \delta G^0{\!}_0 = \sigma^2 \left( \delta T^i{\!}_i - \delta T^0{\!}_0 \right)$, we obtain
\begin{align}
&\dot\kappa + 2H\kappa + \left( 3\dot H + \frac{1}{a^2} \partial_i\partial^i \right) \alpha
 = \frac{\sigma^2}{2} \Big[ 2 \delta\rho_\texttt{R} \nonumber\\
&\quad + 4\left( \dot{\xi} \dot{\delta\xi} - \alpha \dot{\xi}^2 \right) e^{-s\sigma\phi}
 - 2 V^\prime\delta\xi e^{-b\sigma\phi} - 4\alpha \dot{\phi}^2 \Big],
\label{Pert_Einstein_trace}
\end{align}
where we have taken into account that $\delta T^0{\!}_0 = - (\delta\rho_\texttt{R} + \delta\rho_\xi + \delta\rho_\phi)$ and $\delta T^i{\!}_i = 3(\delta p_\texttt{R} + \delta p_\xi + \delta p_\phi)$, and used Eqs.~(\ref{rho_xi_pert})--(\ref{p_phi_pert}) and the relation $\delta p_\texttt{R} = \delta\rho_\texttt{R}/3$.

Finally, the off-diagonal spatial components of the perturbed Einstein equations, $\delta G^i{\!}_j = \sigma^2 \delta T^i{\!}_j$ ($i\neq j$), yield the shear propagation equation
\begin{equation}
\dot\chi + H\chi - \alpha - \varphi = 0,
\label{Pert_Einstein_off}
\end{equation}
since $\delta T^i{\!}_j=\delta^i_j\delta p = 0$ for $i\neq j$.

Equations (\ref{psirad-perturb}), (\ref{xi-perturb_2}), (\ref{rad-perturb_2}), (\ref{Pert_Einstein00}), and (\ref{Pert_Einstein0i})--(\ref{Pert_Einstein_off}) are the equations that, along with the background equations (\ref{ddot-xi-back})--(\ref{dotH}), will be solved numerically in the next section.
They are written in a general gauge, with all four scalar perturbations $\alpha$, $\beta$, $\varphi$, and $\gamma$ left unfixed; the specific gauge choice adopted for the numerical implementation is discussed in Sec.~\ref{sec:numerical}.

\section{Numerical methods\label{sec:numerical}}

We now focus on how to solve the perturbation equations derived in Sec.~\ref{sec:perturbation}.
The procedure follows the stochastic approach implemented in Ref.~\cite{Rodrigues:2025neh}, suitably generalized to the two-scalar-field background introduced in Sec.~\ref{sec:background}.
The numerical computations are performed with \emph{Mathematica}, starting from the publicly available WI2easy package~\cite{Rodrigues:2025neh}, which we adapted to the two-scalar-field cosmological model by introducing the field $\phi$, as described in this section.
We remark that, throughout the numerical analysis, the field $\phi$ is retained as a fully dynamical background degree of freedom, while its perturbation is neglected, $\delta\phi=0$. 
Further, we emphasize that all the background quantities are directly computed as shown in Sec.~\ref{sec:background} and are thus not discussed in the present section.

\subsection{Gauge choice and perturbed Einstein equations}

The perturbation equations obtained in Sec.~\ref{sec:perturbation} are written in a gauge-ready form. 
For the numerical computation, we adopt the uniform-curvature gauge,
\begin{equation}
    \varphi=0.
    \label{uniform_curvature_gauge}
\end{equation} Further, for the sake of convenience, we adopt the Fourier decomposition, implying $\partial_i\partial^i\rightarrow-k^2$, where $k$ denotes the comoving wavenumber.

The numerical evolution is performed in terms of the number of $e$-folds $N$, defined as $dN=Hdt$, so that $a\equiv a(N)$.
This variable differs only by an additive constant from the variable $u$ introduced in Sec.~\ref{sec:background}, specifically $N=u-u_i$, so that their derivatives coincide. 
In particular, the transformation rules yield
\begin{equation}
    \frac{d}{dt}
    =
    H\frac{d}{dN},\quad \frac{d^2}{dt^2}
    =
    H^2\frac{d^2}{dN^2}
    +
    \dot H\frac{d}{dN}.
    \label{second_N_derivative}
\end{equation}

During the inflationary regime relevant for the computation of primordial perturbations, the dissipation coefficient in Eq.~\eqref{gamma} is proportional to $T^p$. 
Its perturbation is therefore
\begin{equation}
    \delta\Gamma_\xi = \frac{\partial \Gamma_\xi}{\partial T} \delta T  =
    \frac{p\Gamma_\xi}{4\rho_\texttt{R}}
    \delta\rho_\texttt{R},
    \label{delta_Gamma_num}
\end{equation}
where we have used $\delta T/T={\delta\rho_\texttt{R}}/({4\rho_\texttt{R}})$.
In our scenario, no contribution proportional to $\delta\xi$ appears in Eq.~\eqref{delta_Gamma_num}, since the dissipation coefficient considered in our two-scalar-field model has no explicit dependence on the inflaton field, but \textit{a priori}, other cases, such as $\Gamma\propto T^p\xi^l$, could also be considered.

Within these choices, the perturbed Einstein equations can be written as follows.

The energy constraint, given by Eq.~\eqref{Pert_Einstein00}, yields
\begin{align}
    \kappa
    =-\frac{\sigma^2}{2H}
    \bigg[
        &e^{-s\sigma\phi}\dot\xi H
        \frac{d\delta\xi}{dN}
        -
        \left(
            e^{-s\sigma\phi}\dot\xi^2
            +
            \dot\phi^2
        \right)\alpha
        \nonumber\\
        &+
        e^{-b\sigma\phi}V^\prime\delta\xi
        +
        \delta\rho_\texttt{R}
    \bigg].
    \label{kappa_uniform}
\end{align}
Combining the momentum constraint, given by Eq.~\eqref{Pert_Einstein0i}, with the definition of $\kappa$, we obtain
\begin{equation}
    \alpha
    =
    \frac{\sigma^2}{2H}
    \left(
        e^{-s\sigma\phi}\dot\xi\,\delta\xi
        -
        \psi_\texttt{R}
    \right).
    \label{alpha_uniform}
\end{equation}
Taking the derivative with respect to $N$ of the above expression for the lapse perturbation $\alpha$ and using the background equation~(\ref{ddot-xi-back}) and the definition of $Q_\xi$ given by Eq.~(\ref{diss-ratio}) yields
\begin{align}
    \frac{d\alpha}{dN}
    ={}&
    -\frac{\dot H}{H^2}\alpha
    -
    \frac{\sigma^2}{2H^2}
    \bigg[
        3H(1+Q_\xi)
        e^{-s\sigma\phi}\dot\xi
        \nonumber\\
        &\hspace{-5mm}
        +
        e^{-b\sigma\phi}V^\prime
    \bigg]\delta\xi
    +
    \frac{\sigma^2}{2H}
    e^{-s\sigma\phi}\dot\xi
    \frac{d\delta\xi}{dN}
-   \frac{\sigma^2}{2H}
    \frac{d\psi_\texttt{R}}{dN}.
    \label{alphaN_uniform}
\end{align}

Finally, the shear $\chi$, which is not required as an independent variable, follows from the definition of $\kappa$, Eq.~(\ref{def-kappa})
\begin{equation}
    \chi
    =
    \frac{a^2}{k^2}
    \left(
        \kappa-3H\alpha
    \right).
    \label{chi_uniform}
\end{equation}\vspace{0mm}

\subsection{Reduced perturbation system}\label{sec:4.2}

Using Eq.~\eqref{second_N_derivative}, the inflaton perturbation equation, Eq.~\eqref{xi-perturb_2}, and the radiation energy-density perturbation equation, Eq.~\eqref{rad-perturb_2}, become
\begin{widetext}
\begin{align}
\frac{d^2\delta\xi}{dN^2} ={}&
-\left[
    \frac{\dot H}{H^2}
    +
    3(1+Q_\xi)
    -
    s\sigma\frac{\dot\phi}{H}
\right]
\frac{d\delta\xi}{dN}
-
\frac{
    k^2/a^2
    +
    e^{(s-b)\sigma\phi}V^{\prime\prime}
}{H^2}
\delta\xi
-
\frac{
    e^{s\sigma\phi}\dot\xi
}{H^2}
\delta\Gamma_\xi
+
\frac{\dot\xi}{H^2}\kappa
+
\frac{\dot\xi}{H}
\frac{d\alpha}{dN}
\nonumber\\
&
-
\frac{
    3H(1+Q_\xi)\dot\xi
    +
    2e^{(s-b)\sigma\phi}V^\prime
}{H^2}
\alpha
+
\zeta_T+\zeta_Q, 
\label{xi_N_num}\\
\frac{d\delta\rho_\texttt{R}}{dN}
={}&
-4\delta\rho_\texttt{R}
+
\frac{\dot\xi^2}{H}
\delta\Gamma_\xi
+
\frac{k^2}{a^2H}\psi_\texttt{R}
+
6HQ_\xi e^{-s\sigma\phi}\dot\xi
\frac{d\delta\xi}{dN}
+
\frac{4\rho_\texttt{R}}{3H}\kappa
+
3
\left(
    -\frac{4}{3}\rho_\texttt{R}
    -
    Q_\xi e^{-s\sigma\phi}\dot\xi^2
\right)\alpha
+
\zeta_\texttt{R},
\label{rhoR_N_num}
\end{align}
\end{widetext}
respectively, while the radiation momentum perturbation satisfies
\begin{equation}
\frac{d\psi_\texttt{R}}{dN}
=
-3\psi_\texttt{R}
-
\frac{\delta\rho_\texttt{R}}{3H}
-
3Q_\xi e^{-s\sigma\phi}\dot\xi\,\delta\xi
-
\frac{4\rho_\texttt{R}}{3H}\alpha.
\label{psiR_N_num}
\end{equation}

We now substitute Eqs.~\eqref{kappa_uniform}--\eqref{alphaN_uniform} into Eqs.~\eqref{xi_N_num}--\eqref{psiR_N_num}, so that the system closes on the four variables
\begin{equation}
    {\boldsymbol{\Phi}}
    =
    \begin{pmatrix}
        \delta\xi\\[1mm]
        d\delta\xi/dN\\[1mm]
        \delta\rho_\texttt{R}\\[1mm]
        \psi_\texttt{R}
    \end{pmatrix}.
    \label{Phi_vector}
\end{equation}

The resulting system can be written in the matrix Langevin form
\begin{equation}
    \frac{d{\boldsymbol{\Phi}}}{dN}
    +
    {\bf A}{\boldsymbol{\Phi}}
    =
    {\bf B}{\boldsymbol{\varsigma}},
    \label{Langevin_matrix}
\end{equation}
where the right-hand side contains the noise contributions.
The matrix ${\bf A}$ has the structure
\begin{equation}
{\bf A}
=
\begin{pmatrix}
0&-1&0&0\\
A_{21}&A_{22}&A_{23}&A_{24}\\
A_{31}&A_{32}&A_{33}&A_{34}\\
A_{41}&0&A_{43}&A_{44}
\end{pmatrix},
\label{A_matrix}
\end{equation}
where the explicit expressions for $A_{ij}$ are given in Appendix~\ref{app:perturbation_matrix}. 

To obtain the right-hand side of Eq.~(\ref{Langevin_matrix}), we  rewrite 
the noise contributions of Eqs.~(\ref{eq:zetaT})--(\ref{eq:zetaR}) in terms of the $e$-fold variable $N$.
Since the inflaton perturbation equation is second order in cosmic time, while the radiation energy-density equation is first order, the noise terms are rescaled as $\zeta_{T,Q}({\bf k},N)=\zeta_{T,Q}({\bf k},t)/H^2$ and $\zeta_\texttt{R}({\bf k},N)=\zeta_\texttt{R}({\bf k},t)/H$, while $\delta(t-t')=H\delta(N-N')$.
We define $\zeta_\xi\equiv\zeta_T+\zeta_Q$ and assume that $\zeta_T$, $\zeta_Q$, and $\zeta_R$ are mutually uncorrelated, so that 
\begin{align}
\left\langle \zeta_A({\bf k},N) \zeta_B({\bf k}',N') \right\rangle
&= B_A^2\delta_{AB} \delta(N-N') (2\pi)^3 \delta({\bf k}+{\bf k}'),
\label{noise_N_corr}
\end{align}
where the subscripts $A$ and $B$ each run over $\xi$ and $\texttt{R}$, and the amplitudes $B_A$ are given below.
The noise vector in Eq.~(\ref{Langevin_matrix}) is then
\begin{equation}
{\boldsymbol{\varsigma}}
    =
    \begin{pmatrix}
        0\\[1mm]
        \varsigma_\xi\\[1mm]
        \varsigma_\texttt{R}\\[1mm]
        0
    \end{pmatrix},
\qquad 
    \varsigma_A({\bf k},N)\equiv\frac{\zeta_A({\bf k},N)}{B_A},
    \label{varsigma_vector}
\end{equation}
whose nonzero components are normalized Gaussian whites noises with zero mean and
\begin{align}
\left\langle \varsigma_A({\bf k},N) \varsigma_B({\bf k}',N') \right\rangle = \delta_{AB} \delta(N-N') (2\pi)^3 \delta({\bf k}+{\bf k}').
\label{noise_N_corr2}
\end{align}
Using Eqs.~(\ref{eq:zetaT})--(\ref{eq:zetaR}), the amplitudes read
\begin{align}
   &B_\xi=\sqrt{B_T^2+B_Q^2}, \\
        &B_\texttt{R}
    =
    -\frac{H\dot\xi}{(Ha)^{3/2}}
    \left(
        6HQ_\xi e^{-s\sigma\phi}T
    \right)^{1/2},
    \label{BR_num}
\end{align}
with
\begin{align}
    &B_T
    =
    \frac{ 1}{(Ha)^{3/2}}
    \left(
        6HQ_\xi T e^{s\sigma\phi}
    \right)^{1/2},
    \label{BT_num}\\
     &B_Q
    =
    \frac{1}{(Ha)^{3/2}}
    \left[
        H^2e^{s\sigma\phi}
        \frac{\sqrt{9+12\pi Q_\xi}}{\pi}
        \left(
            1+2n
        \right)
    \right]^{1/2}.
    \label{BQ_num}
\end{align}
The factors $e^{s\sigma\phi}$ in Eqs.~(\ref{BT_num}) and (\ref{BQ_num}) arise from the noncanonical kinetic coefficient of the inflaton field and from the definition of $Q_\xi$. 
In Eq.~(\ref{BR_num}), the factor $e^{-s\sigma\phi}$ follows from $\Gamma_\xi=3HQ_\xi e^{-s\sigma\phi}$, and the minus sign is adopted by convention, in analogy to Eq.~(\ref{eq:diss_source}).

Thus, the noise matrix entering Eq.~(\ref{Langevin_matrix}) is
\begin{equation}
    {\bf B}
    =
    {\rm diag}
    \left(
        0,B_\xi,B_\texttt{R},0
    \right).
    \label{B_matrix}
\end{equation}
We recall that no independent stochastic source directly acts on the radiation momentum perturbation.

\subsection{Fokker--Planck formulation and initial conditions}

Instead of solving Eq.~\eqref{Langevin_matrix} for a large ensemble of stochastic realizations, we follow the Fokker--Planck approach \cite{Ballesteros:2023dno, Rodrigues:2025neh}. 
Since the Fourier modes are in general complex, we introduce the probability distribution $P(\boldsymbol\Phi,\boldsymbol\Phi^\ast,N)$, which obeys the Fokker–Planck equation associated with Eq.~\eqref{Langevin_matrix},
\begin{align}
\frac{\partial P}{\partial N}
=
\sum_{i,j}
\bigg[
&A_{ij}
\frac{\partial}{\partial\Phi_i}
\left(
    \Phi_jP
\right)
+
A_{ij}
\frac{\partial}{\partial\Phi_i^\ast}
\left(
    \Phi_j^\ast P
\right)
\nonumber\\
&+
\left(
    {\bf B}{\bf B}^{\dagger}
\right)_{ij}
\frac{\partial^2P}
{\partial\Phi_i\partial\Phi_j^\ast}
\bigg].
\label{Fokker_Planck}
\end{align}

The two-point correlation function required to compute the scalar power spectrum is contained in the covariance matrix
\begin{equation}
    {\bf J}
    \equiv
    \left\langle
        {\boldsymbol{\Phi}}
        {\boldsymbol{\Phi}}^\dagger
    \right\rangle.
    \label{J_def}
\end{equation}
Multiplying the Fokker--Planck equation~\eqref{Fokker_Planck} by $\Phi_m\Phi_n^*$, integrating over the perturbation phase space, and integrating by parts yields
\begin{equation}
    \frac{d{\bf J}}{dN}
    =
    -{\bf A}{\bf J}
    -
    {\bf J}{\bf A}^{\dagger}
    +
    {\bf B}{\bf B}^{\dagger}.
    \label{J_evolution}
\end{equation}
This deterministic equation is integrated numerically for each Fourier mode from $N_i=N_\ast-\Delta N$ to $N_f=N_\ast+\Delta N$, with $\Delta N=7$, so that each mode starts well inside the Hubble horizon and ends well outside it. 
Here, $N_\ast$ denotes the time at which the pivot scale $k_\ast$ crosses the Hubble horizon, $k_\ast=a_\ast H_\ast$.
Within this choice, for the thermalized inflaton fluctuations, the covariance matrix is initialized sufficiently deep inside the sub-Hubble regime. 
Indeed, since perturbations originate from stochastic sources during their evolution, we expect that in this regime they are safely assumed to be uncorrelated. 
Thus, the initial state can be set as
\begin{equation}
    {\bf J}(N_i)
    =
    {\bf 0}.
    \label{J_initial}
\end{equation}
The perturbations are consequently generated dynamically by the stochastic terms during the subsequent evolution.

\subsection{Scalar power spectrum and inflationary observables}

In the uniform-curvature gauge (\ref{uniform_curvature_gauge}), the spatial curvature perturbation vanishes, and the comoving curvature perturbation can be written as 
\begin{equation}
    {\cal R}
    =
    -\frac{H}{\rho_{\rm tot}+p_{\rm tot}}\,
    \Psi_{\rm tot}=
    \frac{H}{\rho_{\rm tot}+p_{\rm tot}}
    \left(
        e^{-s\sigma\phi}\dot\xi\,\delta\xi
        -
        \psi_\texttt{R}
    \right),
    \label{R_total_momentum}
\end{equation}
where $\Psi_{\rm tot}$ is the total momentum potential, whose contribution to the comoving curvature perturbation must in general be included in WI \cite{Bastero-Gil:2019rsp}.
Further, the total background enthalpy is introduced directly from the scalar-field and radiation contributions,
\begin{equation}
    \rho_{\rm tot}+p_{\rm tot}
    =
    \dot\phi^2
    +
    e^{-s\sigma\phi}\dot\xi^2
    +
    \frac{4}{3}\rho_\texttt{R}.
    \label{rho_p_num}
\end{equation}
We remark that, within the simplifying assumption used in this paper, i.e., $\delta\phi =0$, the field $\phi$ contributes to the background enthalpy but not directly to the scalar momentum perturbation.

From the Fokker--Planck formulation previously introduced, the dynamical perturbation variables are collected into the vector $\boldsymbol{\Phi}$, while the statistical information about the perturbations is encoded in the covariance matrix ${\bf J}(k,N)$,
whose evolution is determined by Eq.~\eqref{J_evolution}.
Thus, instead of generating an ensemble of individual stochastic realizations, we obtain the two-point correlators required for the power spectrum directly from the deterministic evolution of ${\bf J}$. 
Indeed, within this prescription, Eq.~\eqref{R_total_momentum} can be written as
\begin{equation}
    {\cal R}
    =
    \frac{H}{\rho_{\rm tot}+p_{\rm tot}}\,
    {\bf w}^{\dagger}\boldsymbol{\Phi},
    \qquad
    {\bf w}
    =
    \begin{pmatrix}
        e^{-s\sigma\phi}\dot\xi\\
        0\\
        0\\
        -1
    \end{pmatrix}.
    \label{R_projection}
\end{equation}
We define the dimensionless curvature power spectrum as
\begin{equation}
    P_{\cal R}(k,N)
    =
    \frac{k^3}{2\pi^2}
    \left\langle
        |{\cal R}(k,N)|^2
    \right\rangle,
    \label{PR_definition}
\end{equation}
which, using Eqs.~\eqref{J_def} and \eqref{R_projection}, reads
\begin{equation}
    P_{\cal R}(k,N)
    =
    \frac{k^3}{2\pi^2}
    \frac{H^2}{(\rho_{\rm tot}+p_{\rm tot})^2}
    {\bf w}^{\dagger}
    {\bf J}(k,N)
    {\bf w}.
    \label{PR_direct}
\end{equation}
Notably, although $d\delta\xi/dN$ and $\delta\rho_\texttt{R}$ do not appear explicitly in Eq.~\eqref{PR_direct}, they affect the spectrum indirectly through the coupled evolution of the full covariance matrix.
As a consistency check, in the limit $\dot\phi\rightarrow0,\,e^{-s\sigma\phi}\rightarrow1$, Eqs.~\eqref{R_total_momentum} and \eqref{rho_p_num} reduce to the standard single-field expression.

Finally, to compare our predictions with observational data \cite{Planck:2018jri, ACT:2025tim}, we focus on the pivot scale $k_\ast=0.05\,{\rm Mpc}^{-1}$.
An asterisk on a background quantity, such as $H_\ast$ or $Q_{\xi\ast}$, denotes its value at $N_\ast$.
The scalar amplitude $P_{\cal R}(k_\ast)$, by contrast, is not evaluated at $N_\ast$ but well after Hubble crossing, once the mode has entered the super-Hubble regime and the curvature spectrum has approximately reached its asymptotic value, namely when
\begin{equation}
    \frac{k}{a(N_f)H(N_f)}\ll1.
\end{equation}
For the pivot mode, this condition is satisfied at the final integration time $N_f$, so we define
\begin{equation}
    P_{\cal R}(k_\ast)
    \equiv
    P_{\cal R}(k_\ast,N_f).
\end{equation}

The scalar spectral index $n_s$ is defined by
\begin{equation}
    n_s-1
    =
    \left.
    \frac{d\ln P_{\cal R}(k)}
         {d\ln k}
    \right|_{k=k_\ast},
    \label{ns_definition}
\end{equation}
where the derivative is numerically computed considering a neighborhood of $k_\ast$.

In our approximation, tensor perturbations are not directly affected by the thermal bath at linear order, so we adopt the standard vacuum spectrum
\begin{equation}
    P_T(k_\ast)
    =
    \frac{2\sigma^2H_\ast^2}{\pi^2}.
    \label{PT_numeric}
\end{equation}
The tensor-to-scalar ratio $r$ is then
\begin{equation}
    r
    =
    \frac{P_T(k_\ast)}
         {P_{\cal R}(k_\ast)}
    =
    \frac{
        2\sigma^2H_\ast^2
    }{
        \pi^2P_{\cal R}(k_\ast)
    }.
    \label{r_numeric}
\end{equation}
Thus, all the nontrivial effects associated with the dissipative dynamics, the radiation perturbations, and the noncanonical coupling to the background field $\phi$ enter $r$ only through the numerically computed scalar spectrum $P_{\cal R}(k_\ast)$.

\subsection{Consistency check of the reduced approach \label{subsec-consistency-check}}

In our linear analysis of scalar perturbations in the two-scalar-field WI model, we have neglected both the direct energy transfer from the dark-energy field $\phi$ to radiation and the backreaction of its perturbation $\delta\phi$ on the inflaton and metric perturbations, i.e., we have assumed $\bar{\Gamma}_\phi = 0$ and $\delta\phi=0$. 

In this regard, we assess \emph{a posteriori} whether these assumptions are supported by the dynamics in the region relevant for the pivot scale. 
We start from the background solution and restrict the analysis to the interval $|N-N_*|\leq 1$, with the perturbations evaluated at the pivot mode $k_\ast$.

First, we quantify how much the field $\phi$ is subdominant by considering $\Omega_\phi\equiv{\rho_\phi}/({\rho_\xi+\rho_\phi+\rho_\texttt{R}})$, where all the densities were defined in Sec.~\ref{sec:background}.
For the best-fitting scenario, introduced in that section, we find $\Omega_\phi\sim10^{-5}$ with a small variation over the aforementioned interval. 
This supports the subdominance of the $\phi$ sector at the background level.

Second, we introduce the ratio between the kinetic contributions of the two scalar fields, ${\cal K}_{\phi/\xi}\equiv{\dot\phi^2}/({e^{-s\sigma\phi}\dot\xi^2})$.
In the best-fitting scenario, we obtain ${\cal K}_{\phi/\xi}\sim10^{-2}$, indicating that the kinetic contribution of the field $\phi$ is suppressed with respect to that of the inflaton.

Finally, at the perturbative level, we can study Eq.~\eqref{phi-perturb} by writing it as a residual after imposing $\delta\phi=0$,
\begin{align}
R_{\delta\phi} \equiv\;&
    \dot\phi\left( \dot\alpha+\kappa-3H\alpha
    \right) -s\sigma\dot\xi\,\dot{\delta\xi}\, e^{-s\sigma\phi} \nonumber
\\
    & + b\sigma  V^\prime \,\delta\xi\, e^{-b\sigma\phi}
    +2b\sigma V(\xi)\,\alpha\, e^{-b\sigma\phi}.
    \label{eq:R_generic}
\end{align}
By analogy with the first-order system introduced in Sec.~\ref{sec:4.2}, we can recast the residual as
\begin{equation}
    R_{\delta\phi}(N,k)
    =
    {\bf c}(N)\,{\bf \Phi}(N,k),
    \label{eq:R_delta_phi_vector}
\end{equation}
where ${\bf c}(N) = \begin{pmatrix} c_1 & c_2 & c_3 & c_4 \end{pmatrix}$ with
\begin{align}
c_1 &= -e^{-s\sigma\phi}\,H\frac{d\xi}{dN}\,c_4
      + \left(b\sigma - \sigma^2\frac{d\phi}{dN}\right) V_{,\xi}\, e^{-b\sigma\phi},
    \\
c_2 &= -s\sigma H^2\,\frac{d\xi}{dN}\, e^{-s\sigma\phi},
\\
c_3 &= -\frac{\sigma^2}{3}\frac{d\phi}{dN},
\\
c_4 &= \sigma^2\left[\left(3H+\frac{dH}{dN}\right) \frac{d\phi}{dN} - \frac{b\sigma V}{H}\, e^{-b\sigma\phi}\right].
\end{align}
In the Fokker–Planck formalism, the departure of $R_{\delta\phi}$ from zero is quantified by the amplitude
\begin{equation}
    \mathcal{R}_{\delta\phi}(N,k)\equiv \sqrt{\left\langle
    |R_{\delta\phi}|^2
    \right\rangle}
    =
    \sqrt{
    {\bf c}(N)\,
    {\bf J}(N,k)\,
    {\bf c}^{\dagger}(N)
    }.
    \label{eq:R_delta_phi_rms}
\end{equation}
When numerically evaluated, in Fourier space, it turns out to be $\mathcal{R}_{\delta\phi}(N_\ast,k_\ast)\sim10^{-14} \,m_p^{3/2}$.

The perturbation $\delta\phi$ could affect the observables through its contribution $\psi_\phi=-\dot\phi\,\delta\phi$ to the total momentum potential, which enters the comoving curvature perturbation, Eq.~\eqref{R_total_momentum}, and hence the curvature power spectrum, Eq.~\eqref{PR_direct}.
This contribution is proportional to $\dot\phi$, and, as we have seen, the kinetic energy of $\phi$ is suppressed relative to that of the inflaton, while the smallness of $\mathcal{R}_{\delta\phi}$ suggests that $\delta\phi\ll\delta\xi$. 
These arguments provide an \emph{a posteriori} justification for neglecting $\delta\phi$ as a first step toward a full two-scalar-field generalization of the present work.

\section{Tensor-to-scalar ratio and spectral tilt \label{sec:r_and_ns}}

\begin{figure*}[t]
    \centering
    \includegraphics[scale=1.28]{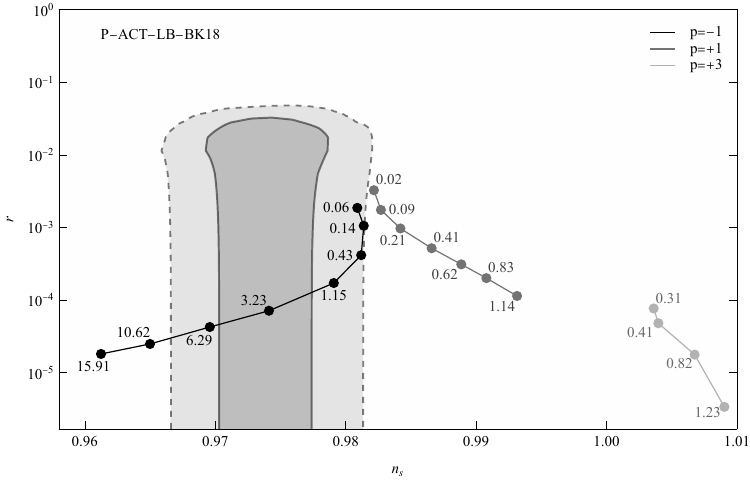}
    \caption{The dark (light) gray region corresponds to the 68\% (95\%) C.L. of the P-ACT-LB-BK18 data combination, which includes Planck, ACT, DESI, and BICEP/Keck data \cite{ACT:2025tim, Planck:2018jri, DESI:2024mwx, BICEP:2021xfz}. 
    The points in the $(n_s, r)$ plane correspond to scenarios with dissipation coefficients $\Gamma_\xi\propto T^{-1}$, $T$, and $T^3$.
    The numerical labels indicate the corresponding values of the dissipation ratio $Q_{\xi\ast}$; the lines connecting the points only guide the eye.}
    \label{fig:rXns}
\end{figure*}

\begin{figure}[t]
    \centering
    \includegraphics[scale=1.3]{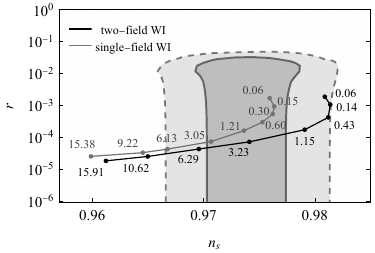}
    \caption{The dark (light) gray region is as in Fig.~\ref{fig:rXns}.
    The points in the $(n_s, r)$ plane compare the predictions of standard single-field WI (gray) and of the two-scalar-field model (black), both with a dissipation coefficient proportional to $T^{-1}$.
    The numerical labels indicate the values of the dissipation ratio at Hubble crossing of the pivot scale, $Q_\ast$ (gray) and $Q_{\xi\ast}$ (black).
   }
    \label{fig:rXns_conf}
\end{figure}

With the scheme developed in Sec.~\ref{sec:numerical}, we are now able to compute observational quantities such as the scalar power spectrum, the tensor-to-scalar ratio, and the spectral tilt, and to confront the two-scalar-field model with the observational bounds \cite{Planck:2018jri, ACT:2025tim}.

We focus on three cases, $p=-1$, $1$, and $3$, for which the dissipation coefficient scales during inflation as $\Gamma_\xi \propto T^{-1}$, $T$ and $T^{3}$, respectively.
In Sec.~\ref{sec:background}, we illustrated the background dynamics with an example for $p=-1$, for which we specified the model parameters and initial conditions.
In all cases considered here, $s$, $b$, $q$, $\phi(u_i)$, and $\phi_u(u_i)$ are kept at the values adopted there, while the remaining parameters, $V_a$, $m$, and $f_\xi$, together with the initial conditions $\xi(u_i)$, $\xi_u(u_i)$, and $\rho_\texttt{R}(u_i)$, are fixed by requiring that (i) the model reproduce the standard cosmological history, including the transition from inflation to the radiation-dominated epoch and the subsequent matter- and dark-energy-dominated eras \cite{Sa:2020fvn}; and (ii) the primordial scalar perturbations satisfy the observational constraint on the amplitude of the scalar power spectrum at the pivot scale, as measured by the Planck mission \cite{Planck:2018jri}, namely $P_{\cal R}(k_\ast) = 2.0989^{+0.0296}_{-0.0292} \times10^{-9}$.

From these requirements, we identify a set of scenarios characterized by different values of $Q_{\xi\ast}$, the dissipation ratio, Eq.~\eqref{diss-ratio}, evaluated at Hubble crossing of the pivot scale, covering both the strong ($Q_{\xi\ast}>1$) and the weak ($Q_{\xi\ast}<1$) dissipative regimes.

However, within the parameter ranges explored and the numerical prescription adopted here, fulfilling both conditions was readily achieved in the strong dissipative regime.
In the weak dissipative regime, by contrast, satisfying them proved considerably more difficult, which restricted the scenarios that we could study to $Q_{\xi\ast}\geq0.06$ for $p=-1$, $Q_{\xi\ast}\geq0.02$ for $p=1$, and $Q_{\xi\ast}\geq0.31$ for $p=3$.
This is partly because, in some regions of parameter space, the transition to the radiation-dominated epoch is not trivially guaranteed \cite{Das:2020lut}, and simultaneously imposing the perturbation constraint further increases the computational effort.
A systematic analysis of the very weak dissipative regime, in all three cases, will require significant modifications to the numerical approach, which we leave for future work.

Figure~\ref{fig:rXns} shows the spectral index $n_s$ and the tensor-to-scalar ratio $r$ predicted for the scenarios obtained with $p=-1$, $1$, and $3$, together with the 68\% and 95\% C.L. regions from the P-ACT-LB-BK18 data combination \cite{ACT:2025tim, Planck:2018jri, DESI:2024mwx, BICEP:2021xfz}.

For $p=-1$, both $n_s$ and $r$ increase as $Q_{\xi\ast}$ decreases, until, for the weakest dissipation considered, the trajectory bends back toward smaller $n_s$ while $r$ keeps growing.
All scenarios with $Q_{\xi\ast}\leq6.29$ are compatible with the data at the 95\% C.L., in both the strong and the weak dissipative regimes; the scenario with $Q_{\xi\ast}=3.23$ lies within the 68\% C.L. region.
For larger dissipation ratios ($Q_{\xi\ast}=10.62$ and $15.91$), the spectrum becomes too red, and the corresponding points fall outside the allowed region.
Across this range, $r$ varies from about $2\times10^{-5}$ in the strong dissipative regime to about $2\times10^{-3}$ for the weakest dissipation considered, comparable to the sensitivity targeted by forthcoming CMB polarization experiments \cite{SimonsObservatory:2018koc, LiteBIRD:2022cnt}.

For a given $Q_{\xi\ast}$, as $p$ increases, the predicted spectrum becomes more blue-tilted, shifting the corresponding points toward larger values of $n_s$.
As a consequence, all scenarios obtained for $p=1$ and $3$ lie outside the observationally allowed region.
Since the corresponding points move away from the allowed region as $Q_{\xi\ast}$ increases, larger dissipation ratios were not considered for these cases.
For $p=1$, the points approach the allowed region as $Q_{\xi\ast}$ decreases; the scenario with the smallest dissipation ratio, $Q_{\xi\ast}=0.02$, lies just outside the 95\% C.L. boundary, suggesting that scenarios with even weaker dissipation may be compatible with observational data.
For $p=3$, by contrast, the spectrum is blue-tilted ($n_s>1$); all scenarios remain far from the allowed region, leaving open the question of whether compatibility with observational data can be achieved in the very weak dissipative regime\footnote{For this temperature dependence, compatibility with data has recently been found in the strong dissipative regime, with \mbox{$Q_\ast\simeq36$}, for a single-field model in which inflation is driven by a quartic potential \cite{Mishra:2026ktu}.}.

Among the three temperature dependences investigated for the two-scalar-field cosmological model, the case $\Gamma_\xi\propto T^{-1}$ is, within the scenarios studied, the most promising one.

Figure~\ref{fig:rXns_conf} compares the predictions of the two-scalar-field model with those of standard single-field WI for $p=-1$.
The single-field predictions were computed with the original WI2easy package~\cite{Rodrigues:2025neh}.
Both sets of predictions follow the same qualitative pattern: $n_s$ and $r$ increase as the dissipation ratio decreases, and, in the weak dissipative regime, the trajectory bends back toward smaller $n_s$ while $r$ keeps growing.
The two-scalar-field predictions are, however, systematically shifted toward larger values of $n_s$, and therefore have a larger spectral index for a given tensor-to-scalar ratio.
As a result, some weakly dissipative scenarios that lie within the 68\% C.L.~region in single-field WI fall outside it in the two-scalar-field model, although they remain within the 95\% C.L.~region; conversely, some strongly dissipative scenarios that are too red in single-field WI become compatible with the data at the 95\% C.L.~in the two-scalar-field model.

\section{Summary and final remarks \label{sec:conclusions}}

In this work, we investigated scalar perturbations in a two-scalar-field cosmological model that unifies warm inflation with the dark sector.
Starting from the most general scalar-perturbed FLRW metric, we derived the linear perturbation equations, including the contribution due to the noncanonical kinetic coupling between the inflaton and dark-energy field and due to stochastic fluctuations for both the inflaton and radiation fluids.
In the numerical analysis, we considered the full background dynamics, but we neglected the perturbations and dissipation of the dark-energy field. 
To support this approximation, we provided an \emph{a posteriori} consistency check in the region relevant for the pivot scale.
The resulting coupled system was solved within a Fokker-Planck formulation.
Finally, we computed the primordial curvature power spectrum $P_{\cal R}$, the scalar spectral index $n_s$ and the tensor-to-scalar ratio $r$ to compare them with observational constraints.

In particular, we considered three temperature dependences of the dissipation coefficient, $\Gamma_\xi\propto T^p$, where $p=-1,\,1$, and $3$. 
The inverse-temperature dissipation coefficient provides promising scenarios in both the weak and strong regimes, consistent with the $68\%$ and $95\%$ C.L. regions obtained from combined Planck and Atacama data.
In contrast, all explored cases with $p=1$ and $3$ lie outside the observationally allowed regions.
The latter exhibits a significantly blue-tilted spectrum, while, with $p=1$, the spectrum approaches the viable region in the weak regime, suggesting that a proper analysis of the very weak regime may provide well-fitting scenarios. 
Lastly, a comparison with standard single-field warm inflation for $p=-1$ shows that the two-scalar-field predictions are systematically shifted toward larger values of $n_s$ for a given $r$.

A full treatment including dark-energy field perturbations, and a dedicated exploration of the very weak dissipative regime, is left for future work.

\appendix

\section{Perturbation matrix}
\label{app:perturbation_matrix}

For completeness, we write here the explicit coefficients of the matrix ${\bf A}$ in Eq.~(\ref{Langevin_matrix}).
To keep the expressions sufficiently compact, we introduce---only in this Appendix---the auxiliary quantities 
\begin{align}
    K &\equiv e^{-s\sigma\phi},
    \\
    U_\xi &\equiv e^{-b\sigma\phi}V^\prime=KF_\xi,
    \\
    F_\xi &\equiv e^{(s-b)\sigma\phi}V^\prime,
    \label{appendix_compact_def}
    \\
    F_{\xi\xi} &\equiv
    e^{(s-b) \sigma\phi} V^{\prime\prime},
    \\
    {\Sigma} &\equiv
    K\dot\xi^2+\dot\phi^2+\frac{4}{3}\rho_\texttt{R}.
\end{align} 

The first row is fixed by the definition $\Phi_2=d\delta\xi/dN$, namely
\begin{equation}
    A_{11}=A_{13}=A_{14}=0,
    \qquad
    A_{12}=-1.
\end{equation}

The coefficients entering the inflaton perturbation equation are
\begin{align}
A_{21}
={}&
\frac{k^2/a^2+F_{\xi\xi}}{H^2}
+
\frac{2\sigma^2U_\xi\dot\xi}{H^3}+
\frac{3\sigma^2K\dot\xi^2}{H^2}
\notag\\
&+
\frac{3\sigma^2KQ_\xi\dot\xi^2}{2H^2}
-
\frac{\sigma^4K\dot\xi^2}{2H^4}{\Sigma},
\label{A21_app}
\\[1mm]
A_{22}
={}&
3(1+Q_\xi)
-
s\sigma\frac{\dot\phi}{H}
-
\frac{\sigma^2}{2H^2}{\Sigma},
\label{A22_app}
\\[1mm]
A_{23}
={}&
\frac{\sigma^2\dot\xi}{3H^3}
+
\frac{3pQ_\xi\dot\xi}
{4H\rho_\texttt{R}},
\label{A23_app}
\\[1mm]
A_{24}
={}&
-\frac{\sigma^2F_\xi}{H^3}
-
\frac{3\sigma^2(2+Q_\xi)\dot\xi}{2H^2}
+
\frac{\sigma^4\dot\xi}{2H^4}{\Sigma}.
\label{A24_app}
\end{align}

For the radiation energy-density perturbation, the corresponding coefficients are
\begin{align}
A_{31}
={}&
\frac{2\sigma^2K\rho_\texttt{R}\dot\xi}{H}
+
\frac{3\sigma^2K^2Q_\xi\dot\xi^3}{2H}
+
\frac{2\sigma^2\rho_\texttt{R}U_\xi}{3H^2}\notag\\
&-
\frac{\sigma^4K\rho_\texttt{R}\dot\xi}{3H^3}
\left(
    K\dot\xi^2+\dot\phi^2
\right),
\label{A31_app}
\\[1mm]
A_{32}
={}&
-6HQ_\xi K\dot\xi
+
\frac{2\sigma^2K\rho_\texttt{R}\dot\xi}{3H},
\label{A32_app}
\\[1mm]
A_{33}
={}&
4
+
\frac{2\sigma^2\rho_\texttt{R}}{3H^2}
-
\frac{3pQ_\xi K\dot\xi^2}
{4\rho_\texttt{R}},
\label{A33_app}
\\[1mm]
A_{34}
={}&
-\frac{k^2}{a^2H}
-\frac{2\sigma^2\rho_\texttt{R}}{H}
-\frac{3\sigma^2KQ_\xi\dot\xi^2}{2H}
\nonumber \\
&+ \frac{\sigma^4\rho_\texttt{R}}{3H^3}
\left(
    K\dot\xi^2+\dot\phi^2
\right).
\label{A34_app}
\end{align}

Finally, the radiation momentum equation gives
\begin{align}
A_{41}
&=
3Q_\xi K\dot\xi
+
\frac{2\sigma^2K\rho_\texttt{R}\dot\xi}
{3H^2},
\label{A41_app}
\\
A_{42}
&=0,
\label{A42_app}
\\
A_{43}
&=
\frac{1}{3H},
\label{A43_app}
\\
A_{44}
&=
3-
\frac{2\sigma^2\rho_\texttt{R}}
{3H^2}.
\label{A44_app}
\end{align}

In the limit $s\rightarrow0$, $\dot\phi\rightarrow0$, and hence $K\rightarrow1$, the above expressions reduce to the corresponding single-field coefficients \cite{Rodrigues:2025neh}. 

\vfill


\begin{thebibliography}{0}

\bibitem{Starobinsky:1980te} A. A. Starobinsky, A new type of isotropic cosmological models without singularity, Phys. Lett. B \textbf{91}, 99 (1980).

\bibitem{Guth:1980zm} A. H. Guth, Inflationary universe: A possible solution to the horizon and flatness problems, Phys. Rev. D \textbf{23}, 347 (1981).

\bibitem{Linde:1981mu} A. D. Linde, A new inflationary universe scenario: A possible solution of the horizon, flatness, homogeneity, isotropy and primordial monopole problems, Phys. Lett. B \textbf{108}, 389 (1982).

\bibitem{Albrecht:1982wi} A. Albrecht and P. J. Steinhardt, Cosmology for grand unified theories with radiatively induced symmetry breaking, Phys. Rev. Lett. \textbf{48}, 1220 (1982).

\bibitem{Linde:1983gd} A. D. Linde, Chaotic inflation, Phys. Lett. B \textbf{129}, 177 (1983).

\bibitem{Mukhanov:1981xt} V. F. Mukhanov and G. V. Chibisov, Quantum fluctuations and a nonsingular universe, JETP Lett. \textbf{33}, 532 (1981).

\bibitem{Starobinsky:1982ee} A. A. Starobinsky, Dynamics of phase transition in the new inflationary universe scenario and generation of perturbations, Phys. Lett. B \textbf{117}, 175 (1982).

\bibitem{Guth:1982ec} A. H. Guth and S.-Y. Pi, Fluctuations in the new inflationary universe, Phys. Rev. Lett. \textbf{49}, 1110 (1982).

\bibitem{Bardeen:1983qw} J. M. Bardeen, P. J. Steinhardt, and M. S. Turner, Spontaneous creation of almost scale-free density perturbations in an inflationary universe, Phys. Rev. D \textbf{28}, 679 (1983).

\bibitem{Starobinsky:1979ty} A. A. Starobinsky, Spectrum of relict gravitational radiation and the early state of the universe, JETP Lett. \textbf{30}, 682 (1979).

\bibitem{Planck:2018jri} Y. Akrami \textit{et al.} (Planck Collaboration), Planck 2018 results. X. Constraints on inflation, Astron. Astrophys. \textbf{641}, A10 (2020).

\bibitem{ACT:2025fju} T. Louis \textit{et al.} (ACT Collaboration), The Atacama Cosmology Telescope: DR6 Power Spectra, Likelihoods and $\Lambda$CDM Parameters, J. Cosmol. Astropart. Phys. 11 (2025) 062.

\bibitem{ACT:2025tim} E. Calabrese \textit{et al.} (ACT Collaboration), The Atacama Cosmology Telescope: DR6 Constraints on Extended Cosmological Models, J. Cosmol. Astropart. Phys. 11 (2025) 063.  

\bibitem{BICEP:2021xfz} P. A. R. Ade \textit{et al.} (BICEP/Keck Collaboration), Improved constraints on primordial gravitational waves using Planck, WMAP, and BICEP/Keck observations through the 2018 observing season, Phys. Rev. Lett. \textbf{127}, 151301 (2021).

\bibitem{Tristram:2021tvh} M. Tristram \textit{et al.}, Improved limits on the tensor-to-scalar ratio using BICEP and Planck data, Phys. Rev. D \textbf{105}, 083524 (2022).

\bibitem{Berera:1995ie} A. Berera, Warm inflation, Phys. Rev. Lett. \textbf{75}, 3218 (1995).

\bibitem{Berera:2008ar} A. Berera, I. G. Moss, and R. O. Ramos, Warm inflation and its microphysical basis, Rep. Prog. Phys. \textbf{72}, 026901 (2009).

\bibitem{Kamali:2023lzq} V. Kamali, M. Motaharfar, and R. O. Ramos, Recent developments in warm inflation, Universe \textbf{9}, 124 (2023).

\bibitem{Berera:2023liv} A. Berera, The warm inflation story, Universe \textbf{9}, 272 (2023).

\bibitem{Berera:2002sp} A. Berera and R. O. Ramos, Construction of a robust warm inflation mechanism, Phys. Lett. B \textbf{567}, 294 (2003).

\bibitem{Bastero-Gil:2010dgy} M. Bastero-Gil, A. Berera, and R. O. Ramos, Dissipation coefficients from scalar and fermion quantum field interactions, J. Cosmol. Astropart. Phys. 09 (2011) 033.

\bibitem{Bastero-Gil:2012akf} M. Bastero-Gil, A. Berera, R. O. Ramos, and J. G. Rosa, General dissipation coefficient in low-temperature warm inflation, J. Cosmol. Astropart. Phys. 01 (2013) 016.

\bibitem{Bastero-Gil:2016qru} M. Bastero-Gil, A. Berera, R. O. Ramos, and J. G. Rosa, Warm Little Inflaton, Phys. Rev. Lett. \textbf{117}, 151301 (2016).

\bibitem{Rosa:2018iwu} J. G. Rosa and L. B. Ventura, Warm Little Inflaton becomes cold dark matter, Phys. Rev. Lett. \textbf{122}, 161301 (2019).

\bibitem{Bastero-Gil:2019gao} M. Bastero-Gil, A. Berera, R. O. Ramos, and J. G. Rosa, Towards a reliable effective field theory of inflation, Phys. Lett. B \textbf{813}, 136055 (2021).

\bibitem{Berghaus:2019whh} K. V. Berghaus, P. W. Graham, and D. E. Kaplan, Minimal warm inflation, J. Cosmol. Astropart. Phys. 03 (2020) 034; \textbf{10} (2023) E02(E).

\bibitem{Bartrum:2013fia} S. Bartrum, M. Bastero-Gil, A. Berera, R. Cerezo, R. O. Ramos, and J. G. Rosa, The importance of being warm (during inflation), Phys. Lett. B \textbf{732}, 116 (2014).

\bibitem{Benetti:2016jhf} M. Benetti and R. O. Ramos, Warm inflation dissipative effects: predictions and constraints from the Planck data, Phys. Rev. D \textbf{95}, 023517 (2017).

\bibitem{Bastero-Gil:2017wwl} M. Bastero-Gil, S. Bhattacharya, K. Dutta, and M. R. Gangopadhyay, Constraining warm inflation with CMB data, J. Cosmol. Astropart. Phys. 02 (2018) 054.

\bibitem{Montefalcone:2023pvh} G. Montefalcone, V. Aragam, L. Visinelli, and K. Freese, WarmSPy: a numerical study of cosmological perturbations in warm inflation, J. Cosmol. Astropart. Phys. 01 (2024) 032.

\bibitem{Berera:2025vsu} A. Berera, S. Brahma, Z. Qiu, R. O. Ramos, and G. S. Rodrigues, The early universe is ACT-ing warm, J. Cosmol. Astropart. Phys. 11 (2025) 059.  

\bibitem{Ballesteros:2023dno} G. Ballesteros, A. P\'erez Rodr\'iguez, and M. Pierre, Monomial warm inflation revisited, J. Cosmol. Astropart. Phys. 03 (2024) 003.

\bibitem{Motaharfar:2018zyb} M. Motaharfar, V. Kamali, and R. O. Ramos, Warm inflation as a way out of the swampland, Phys. Rev. D \textbf{99}, 063513 (2019).

\bibitem{Das:2018hqy} S. Das, Warm inflation in the light of swampland criteria, Phys. Rev. D \textbf{99}, 063514 (2019).

\bibitem{Bertolami:2022knd} O. Bertolami and P. M. S\'a, Multi-field cold and warm inflation and the de Sitter swampland conjectures, J. Cosmol. Astropart. Phys. 09 (2022) 001.

\bibitem{SimonsObservatory:2018koc} P. Ade \textit{et al.} (Simons Observatory Collaboration), The Simons Observatory: science goals and forecasts, J. Cosmol. Astropart. Phys. 02 (2019) 056.

\bibitem{LiteBIRD:2022cnt} E. Allys \textit{et al.} (LiteBIRD Collaboration), Probing cosmic inflation with the LiteBIRD cosmic microwave background polarization survey, Prog. Theor. Exp. Phys. \textbf{2023}, 042F01 (2023).

\bibitem{Riess:1998cb} A. G. Riess \textit{et al.}, Observational evidence from supernovae for an accelerating universe and a cosmological constant, Astron. J. \textbf{116}, 1009 (1998).

\bibitem{Perlmutter:1998np} S. Perlmutter \textit{et al.}, Measurements of $\Omega$ and $\Lambda$ from 42 high-redshift supernovae, Astrophys. J. \textbf{517}, 565 (1999).

\bibitem{Weinberg:1988cp} S. Weinberg, The cosmological constant problem, Rev. Mod. Phys. \textbf{61}, 1 (1989).

\bibitem{DESI:2024mwx} A. G. Adame \textit{et al.} (DESI Collaboration), DESI 2024 VI: Cosmological constraints from the measurements of baryon acoustic oscillations, J. Cosmol. Astropart. Phys. 02 (2025) 021.

\bibitem{DESI:2025zgx} M. Abdul Karim \textit{et al.} (DESI Collaboration), DESI DR2 results. II. Measurements of baryon acoustic oscillations and cosmological constraints, Phys. Rev. D \textbf{112}, 083515 (2025).

\bibitem{Ratra:1987rm} B. Ratra and P. J. E. Peebles, Cosmological consequences of a rolling homogeneous scalar field, Phys. Rev. D \textbf{37}, 3406 (1988).

\bibitem{Wetterich:1987fm} C. Wetterich, Cosmology and the fate of dilatation symmetry, Nucl. Phys. B \textbf{302}, 668 (1988).

\bibitem{Caldwell:1997ii} R. R. Caldwell, R. Dave, and P. J. Steinhardt, Cosmological imprint of an energy component with general equation of state, Phys. Rev. Lett. \textbf{80}, 1582 (1998).

\bibitem{Turner:1983he} M.~S.~Turner, Coherent scalar-field oscillations in an expanding Universe, Phys.\ Rev.\ D \textbf{28}, 1243 (1983).

\bibitem{Peebles:1998qn} P. J. E. Peebles and A. Vilenkin, Quintessential inflation, Phys. Rev. D \textbf{59}, 063505 (1999).

\bibitem{Panotopoulos:2007ri} G. Panotopoulos, How to unify dark matter, dark energy, and inflation, Phys. Rev. D \textbf{75}, 127301 (2007).

\bibitem{Liddle:2008bm} A. R. Liddle, C. Pahud, and L. A. Ure\~na-L\'opez, Triple unification of inflation, dark matter, and dark energy using a single field, Phys. Rev. D \textbf{77}, 121301(R) (2008).

\bibitem{Henriques:2009hv} A. B. Henriques, R. Potting, and P. M. S\'a, Unification of inflation, dark energy, and dark matter within the Salam-Sezgin cosmological model, Phys. Rev. D \textbf{79}, 103522 (2009).

\bibitem{Paliathanasis:2025mvy}
A. Paliathanasis, T. Mengoni, G. Leon and O. Luongo, Constraints on Chiral-Quintom dark energy after DESI DR2 and impact on unifying dark energy with inflation, Fortschr. Phys. (in press), arXiv:2512.00558.

\bibitem{Sa:2020fvn} P.~M.~S\'a, Triple unification of inflation, dark energy, and dark matter in two-scalar-field cosmology, Phys.\ Rev.\ D \textbf{102}, 103519 (2020).

\bibitem{Zhang:2021ppy} X.-M. Zhang, K. Li, Y.-F. Guo, P.-C. Chu, H. Liu, and J.-Y. Zhu, Two models unifying warm inflation with dark matter and dark energy, Phys. Rev. D \textbf{104}, 103513 (2021).

\bibitem{Mishra:2026ktu} S. S. Mishra, U. Kumar, S. Das, and V. Sahni, Dark matter as an inflationary relic in warm inflation, arXiv:2606.22147.

\bibitem{Amendola:1999er} L.~Amendola, Coupled quintessence, Phys.\ Rev.\ D \textbf{62}, 043511 (2000).

\bibitem{Sa:2021eft} P.~M.~S\'a, Late-time evolution of the Universe within a two-scalar-field cosmological model, Phys.\ Rev.\ D \textbf{103}, 123517 (2021).

\bibitem{DAgostino:2021vvv} R.~D'Agostino and O.~Luongo, Cosmological viability of a double field unified model from warm inflation, Phys.\ Lett.\ B \textbf{829}, 137070 (2022).

\bibitem{Henriques:1993km} A. B. Henriques, Graviton creation in an inflationary universe, Phys. Rev. D \textbf{49}, 1771 (1994).

\bibitem{Moorhouse:1994nc} R. G. Moorhouse, A. B. Henriques, and L. E. Mendes, Graviton creation in an expanding universe, Phys. Rev. D \textbf{50}, 2600 (1994).

\bibitem{Mendes:1994ai} L. E. Mendes, A. B. Henriques, and R. G. Moorhouse, Exact calculation of the energy density of cosmological gravitational waves, Phys. Rev. D \textbf{52}, 2083 (1995).

\bibitem{Sa:2007pc} P. M. S\'a and A. B. Henriques, Parametric resonance and cosmological gravitational waves, Phys. Rev. D \textbf{77}, 064002 (2008).

\bibitem{Sa:2008yq} P. M. S\'a and A. B. Henriques, Gravitational wave generation in power-law inflationary models, Gen. Relativ. Gravit. \textbf{41}, 2345 (2009).

\bibitem{Sa:2010qw} P. M. S\'a and A. B. Henriques, Gravitational-wave generation in hybrid quintessential inflationary models, Phys. Rev. D \textbf{81}, 124043 (2010).

\bibitem{Sa:2011rm} P. M. S\'a and A. B. Henriques, Gravitational wave generation in loop quantum cosmology, Phys. Rev. D \textbf{85}, 024034 (2012).

\bibitem{Bouhmadi-Lopez:2009uca} M. Bouhmadi-L\'opez, P. Fraz\~ao, and A. B. Henriques, Stochastic gravitational waves from a new type of modified Chaplygin gas, Phys. Rev. D \textbf{81}, 063504 (2010).

\bibitem{Bouhmadi-Lopez:2012piq} M. Bouhmadi-L\'opez, J. Morais, and A. B. Henriques, Smoking guns of a bounce in modified theories of gravity through the spectrum of gravitational waves, Phys. Rev. D \textbf{87}, 103528 (2013).

\bibitem{Luongo:2025ovo} O. Luongo, T. Mengoni, and P. M. S\'a, Gravitational waves from two scalar fields unifying the dark sector with inflation, Phys. Rev. D \textbf{113}, 023541 (2026).

\bibitem{Luongo:2026wdr} O. Luongo, T. Mengoni, and P. M. S\'a, Gravitational waves from warm inflation in the weak dissipative regime, Phys. Dark Univ. \textbf{52}, 102317 (2026).

\bibitem{Mengoni:2026bog} T. Mengoni, Continuous Bogoliubov formalism for gravitational-wave generation in a unified dark sector warm inflation, Int. J. Mod. Phys. A \textbf{41}, 2646006 (2026).

\bibitem{Wang:2018cev} Y.-Y. Wang, J.-Y. Zhu, and X.-M. Zhang, Two-field warm inflation and its scalar perturbations on large scales, Phys. Rev. D \textbf{97}, 063510 (2018).

\bibitem{Li:2019zbk} X.-B. Li, X.-G. Zheng, and J.-Y. Zhu, Spectra and entropy of multifield warm inflation, Phys. Rev. D \textbf{99}, 043528 (2019).

\bibitem{Wang:2019ozs} Y.-Y. Wang, J.-Y. Zhu, and X.-M. Zhang, Observational constraints on two-field warm inflation, Phys. Rev. D \textbf{99}, 103529 (2019).

\bibitem{Berera:2025jbj} A. Berera, H. Bernardo, S. Brahma, J. Calder\'on-Figueroa, R. O. Ramos, and M. W. Toomey, Heterotic warm inflation, J. High Energy Phys. 06 (2026) 099.

\bibitem{Fixsen:2009ug} D.~J.~Fixsen, The temperature of the cosmic microwave background, Astrophys.\ J.\ \textbf{707}, 916 (2009).

\bibitem{Planck:2018vyg} N.~Aghanim et al. (Planck Collaboration), Planck 2018 results. VI. Cosmological parameters, Astron.\ Astrophys.\ \textbf{641}, A6 (2020).

\bibitem{Gumjudpai:2005} B.~Gumjudpai, T.~Naskar, M.~Sami, and S.~Tsujikawa, Coupled dark energy: Towards a general description of the dynamics, J.\ Cosmol.\ Astropart.\ Phys.\ \textbf{06} (2005) 007.

\bibitem{Boehmer:2009roi} C.~G. B\"{o}hmer, G.~Caldera-Cabral, R.~Lazkoz, and R.~Maartens, Dynamics of dark energy with a coupling to dark matter, Phys.\ Rev.\ D \textbf{78}, 023505 (2008).

\bibitem{Amendola:2014kwa} L.~Amendola, T.~Barreiro, and N.~J.~Nunes, Multifield coupled quintessence, Phys.\ Rev.\ D \textbf{90}, 083508 (2014).

\bibitem{Bahamonde:2017ize} S.~Bahamonde, C.~G.~B\"{o}hmer, S.~Carloni, E.~Copeland, W.~Fang, and N.~Tamanini, Dynamical systems applied to cosmology: Dark energy and modified gravity, Phys.\ Rep.\ \textbf{775-777}, 1 (2018).

\bibitem{Potting:2022} R.~Potting and P.~M.~S\'a, Coupled quintessence with a generalized interaction term, Int.\ J.\ Mod.\ Phys.\ D \textbf{31}, 2250111 (2022).

\bibitem{Graham:2009bf} C.~Graham and I.~G.~Moss, Density fluctuations from warm inflation, J.\ Cosmol.\ Astropart.\ Phys. \textbf{07} (2009) 013.

\bibitem{Bastero-Gil:2011rva} M.~Bastero-Gil, A.~Berera and R.~O.~Ramos, Shear viscous effects on the primordial power spectrum from warm inflation, J.\ Cosmol.\ Astropart.\ Phys. \textbf{07} (2011) 030.

\bibitem{Hwang:2001fb} J.-c.~Hwang and H.~Noh, Cosmological perturbations with multiple fluids and fields, Class. Quant. Grav. \textbf{19}, 527 (2002).

\bibitem{Malik:2002jb} K.~A.~Malik, D.~Wands and C.~Ungarelli, Large-scale curvature and entropy perturbations for multiple interacting fluids, Phys. Rev. D \textbf{67}, 063516 (2003).

\bibitem{Faraoni:2012hn} V.~Faraoni, Correspondence between a scalar field and an effective perfect fluid, Phys.\ Rev.\ D \textbf{85}, 024040 (2012).

\bibitem{Harko:2021gnz} T.~Harko and H.~Sheikhahmadi, Warm inflation with non-comoving scalar field and radiation fluid, Eur.\ Phys.\ J.\ C \textbf{81}, 165 (2021).

\bibitem{Ramos:2013nsa} R. O. Ramos and L. A. da Silva, Power spectrum for inflation models with quantum and thermal noises, J.\ Cosmol.\ Astropart.\ Phys.\ \textbf{03} (2013) 032.

\bibitem{Rodrigues:2025neh}
G.~S.~Rodrigues and R.~O.~Ramos, WI2easy: warm inflation dynamics made easy, J.\ Cosmol.\ Astropart.\ Phys.\ \textbf{09} (2025) 014.

\bibitem{Bastero-Gil:2019rsp} M.~Bastero-Gil, A.~Berera and J.~R.~Calder{\'o}n, Reexamination of the warm inflation curvature perturbations spectrum, J.\ Cosmol.\ Astropart.\ Phys.\ \textbf{07} (2019) 019.

\bibitem{Das:2020lut} S.~Das and R.~O.~Ramos, Graceful exit problem in warm inflation, Phys. Rev. D \textbf{103}, 123520 (2021).

\end{thebibliography}
\end{document}